\documentclass[fleqn,usenatbib]{mnras}

\usepackage{amsfonts}

\usepackage[T1]{fontenc}
\usepackage{ae,aecompl}
\usepackage[position=top]{subfig}
\usepackage{float}
\usepackage{comment}

\usepackage{grffile}
\usepackage{etoolbox}
\makeatletter 
  \patchcmd{\NAT@citex}
    {\@citea\NAT@hyper@{%
      \NAT@nmfmt{\NAT@nm}%
      \hyper@natlinkbreak{\NAT@aysep\NAT@spacechar}{\@citeb\@extra@b@citeb}%
      \NAT@date}}
    {\@citea\NAT@nmfmt{\NAT@nm}%
    \NAT@aysep\NAT@spacechar\NAT@hyper@{\NAT@date}}{}{}

  \patchcmd{\NAT@citex}
    {\@citea\NAT@hyper@{%
      \NAT@nmfmt{\NAT@nm}%
      \hyper@natlinkbreak{\NAT@spacechar\NAT@@open\if*#1*\else#1\NAT@spacechar\fi}%
        {\@citeb\@extra@b@citeb}%
      \NAT@date}}
    {\@citea\NAT@nmfmt{\NAT@nm}%
    \NAT@spacechar\NAT@@open\if*#1*\else#1\NAT@spacechar\fi\NAT@hyper@{\NAT@date}}
    {}{}
\makeatother

\usepackage{etoolbox}
\makeatletter
\patchcmd\@combinedblfloats{\box\@outputbox}{\unvbox\@outputbox}{}{%
   \errmessage{\noexpand\@combinedblfloats could not be patched}%
}%
 \makeatother
\usepackage{graphicx}	
\usepackage{amsmath}	
\usepackage{amssymb}	

\title[BDMS in early structure formation]{Constraining the non-gravitational scattering of baryons and dark matter with early cosmic structure formation}

\author[B. Liu, A. Schauer, V. Bromm]{Boyuan Liu\textsuperscript{\href{https://orcid.org/0000-0002-4966-7450}{\includegraphics[width=2.5mm]{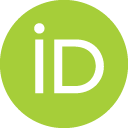}}\,}\thanks{E-mail: boyuan@utexas.edu}$^{1}$, Anna
T. P. Schauer\thanks{Hubble Fellow}$^{1}$, and Volker Bromm$^{1}$
\\
$^{1}$Department of Astronomy, University of Texas, Austin, TX 78712, USA\\
}

\date{Accepted XXX. Received YYY; in original form ZZZ}

\pubyear{2019}

\begin{document}
\label{firstpage}
\pagerange{\pageref{firstpage}--\pageref{lastpage}}
\maketitle

\begin{abstract}
We derive new constraints on the non-gravitational baryon-dark-matter scattering (BDMS) by evaluating the mass thresholds of dark matter (DM) haloes in which primordial gas can cool efficiently to form Population III (Pop~III) stars, based on the timing of the observed 21-cm absorption signal. We focus on the BDMS model with interaction cross-section $\sigma=\sigma_{1}[v/(1\ \mathrm{km\ s^{-1}})]^{-4}$, where $v$ is the relative velocity of the encounter. Our results rule out the region in parameter space with $\sigma_{1}\gtrsim 10^{-19}\ \mathrm{cm^{2}}$ and DM particle mass $m_{\chi}c^{2}\lesssim 3\times 10^{-2}$~GeV, where the cosmic number density of Pop~III hosts at redshift $z\sim 20$ is at least three orders of magnitude smaller than in the standard Lambda cold DM ($\Lambda$CDM) case. In these BDMS models, the formation of Pop~III stars is significantly suppressed for $z\gtrsim 20$, inconsistent with the timing of the observed global 21-cm absorption signal. For the fiducial BDMS model with $m_{\chi}c^{2}=0.3$~GeV and $\sigma_{1}=8\times 10^{-20}\ \mathrm{cm^{2}}$, capable of accommodating the measured absorption depth, the number density of Pop~III hosts is reduced by a factor of $3-10$ at $z\sim 15-20$, when the 21-cm signal is imprinted, compared with the $\Lambda$CDM model. The confluence of future detailed cosmological simulations with improved 21-cm observations promises to probe the particle-physics nature of DM at the small-scale frontier of early structure formation.
\end{abstract}

\begin{keywords}
early universe -- dark ages, reionization, first stars -- dark matter 
\end{keywords}



\section{Introduction}
Recently, the Experiment to Detect the Global Epoch of Reionization Signature (EDGES) measured the 21-cm absorption signal from primordial neutral hydrogen at redshift $z\sim 17$, which is ($3\sigma$) stronger than what is expected from the standard Lambda cold dark matter ($\Lambda$CDM) model \citep{nature}. This signal with its specific timing and strength, if confirmed, contains valuable information on the state of the early Universe, the nature of dark matter (DM), and even new physics. 

The timing of the EDGES signal shows that sufficient star formation has occurred before $z=20$ to establish the Lyman-$\alpha$ radiation field that couples the spin temperature of neutral hydrogen with the kinetic temperature of gas via the Wouthuysen-Field effect \citep{w52, field58}. Recent studies by \citet{madau18} and \citet{Anna2019} show that Population III (Pop~III) stars formed in minihaloes play an important role in this process. Besides, several groups also use this timing information to constrain warm DM (WDM) properties and find that the mass of thermal WDM is limited to $m_{\chi}c^{2}\gtrsim 2-6.1$~keV \citep{Schneider2018,Sitwell2014,Safar2018}. This unique absorption signal can also put constraints on interacting dark matter models \citep{lopez2019dark}.

To interpret the strength of the EDGES signal, \citet{nature1} argues that the intergalactic medium (IGM) at $z\sim 17$ has to be cooler than what is implied by current theoretical predictions, which could be achieved by non-gravitational scattering between baryons and DM particles, such as predicted for millicharged atomic DM \citep{millicharge2012,munoz2018insights}, or axion-like DM \citep{yang2018natural}. An alternative interpretation posits a possible early radio background, in addition to the cosmic microwave background (CMB, \citealt{feng2018enhanced}), which itself may be produced by DM \citep{fraser2018edges}. The former interpretation, if verified, has great significance for fundamental physics. However, the effect of baryon-dark-matter scattering (BDMS) so far has only been explored in the linear regime for the IGM (e.g., \citealt{tashiro2014effects,dvorkin2014constraining,fialkov2018,slatyer2018early}), and a large region in the phenomenological parameter space for such BDMS is consistent with the EDGES signal (see Figure~3 of \citealt{nature1}). When considering the particle physics mechanisms behind BDMS, tight constraints are derived from a variety of probes, such as the CMB, primordial chemistry,  Lyman-$\alpha$ forest power spectra, and laboratory experiments (e.g., \citealt{kovetz2018tighter,berlin2018severely,barkana2018signs,mahdawi2018constraints,Xu2018}). 

The effect of BDMS in the nonlinear regime of high-$z$ structure formation is also important for the 21-cm signature, as the feedback from the first generation of (Pop~III) stars is influencing the thermal state of the early IGM. In this regard, \citet{Hirano2018} investigate the BDMS effect in primordial star-forming gas clouds, reaching central hydrogen number densities of $n_{c}>10^{6}\ \mathrm{cm^{-3}}$. Their study shows that massive stars are able to form within the fiducial BDMS model that can accommodate the absorption depth measured by EDGES, and rules out a large region in parameter space, where strong cooling or heating can occur to suppress formation of massive Pop~III stars. However, their work is not fully self-consistent, since the physical properties of star-forming clouds in virialized haloes are derived from simulations carried out without the BDMS effect, and the momentum transfer as well as the thermal back-reaction on the DM fluid by BDMS is ignored. 

In general, the abundance and states of star-forming clouds can be quite different with BDMS, compared with the $\Lambda$CDM baseline model. Thus, it is still unknown whether BDMS can self-consistently accommodate the EDGES signal, including its timing and strength, and what constraints on the BDMS parameter space can be obtained from early structure formation. In this study, we address these questions by evaluating the mass thresholds of DM haloes that can host Pop~III stars with BDMS. To be more specific, we focus on the pre-virialization stage of potential star-forming clouds, and calculate the thermal and chemical histories of DM haloes in the presence of BDMS, to examine whether the gas inside them can efficiently cool to form stars. 

This paper is structured as follows. In Section~\ref{s2}, we describe the one-zone model used to calculate the mass thresholds of haloes that can efficiently cool to host Pop~III stars, when BDMS is included. In Section~\ref{s3}, we present the results for the fiducial BDMS model (Sec.~\ref{s3.1}), as well as the constraints on the wider BDMS parameter space (Sec.~\ref{s3.2}). Finally, our conclusions and perspectives for future studies are discussed in Section~\ref{s4}.

\section{Methodology}
\label{s2}
Following \citet{BDMS}, throughout this work we assume that DM can be regarded as an ideal gas (with a Maxwell-Boltzmann velocity distribution) of thermal temperature $T_{\chi}$ and adiabatic index $\gamma_{\chi}$. 
Note that this ideal gas assumption only applies in the limit where DM self-interactions are frequent enough to efficiently redistribute DM velocities, and it can overestimate the energy-transfer rate by up to a factor of 3 in the early coupling scenarios where DM starts thermally and kinematically coupled to baryons \citep{ali2019boltzmann}. However, here we focus on a late coupling scenario in which the BDMS (momentum-transfer) cross-section is parametrized as $\sigma=\sigma_{0}v^{-4}=\sigma_{1}[v/(1\ \mathrm{km\ s^{-1}})]^{-4}$, where $v$ is the relative velocity of the encounter. Such a strong inverse velocity dependence would arise naturally in Coulomb-like scattering. It is also necessary to produce sufficient cooling in the cosmic dark ages ($20\lesssim z\lesssim 200$), when the relative velocities between DM and baryons are at their minimum value, to account for the 21-cm absorption signal, while remaining consistent with observations such as the CMB and Lyman-$\alpha$ forest (see \citealt{dvorkin2014constraining,Xu2018} for general models with $\sigma\propto v^{n}$, $-4\le n \le 2$). It has previously been shown that such scattering with $\sigma\propto v^{-4}$ is indeed able to explain the strength of the EDGES signal \citep{nature1,slatyer2018early}. Unfortunately, going beyond the ideal gas assumption for late-coupling self-interacting DM is non-trivial, where solving the exact collisional Boltzmann equation seems the only approach \citep{ali2019boltzmann}. This is beyond the scope of our study. Any uncertainties resulting from this approximation are expected to be within factors of a few. 

Under the ideal gas assumption and $\sigma\propto v^{-4}$ parameterization, we work in a phenomenological framework, where the properties of BDMS are determined by two parameters: DM particle mass $m_{\chi}$ and cross-section parameter $\sigma_{1}$. The constraints derived in this $m_{\chi}$-$\sigma_{1}$ parameter space are general and applicable for any particle physics models with Coulomb-like BDMS. For example, such Coulomb-like interactions exists in millicharged DM models, where the dark particles carry a small electric charge, as in the hidden Stueckelberg Z' model with millicharged fermion-antifermion pairs \citep{cheung2007hidden}, or in the atomic DM scenario with small charge induced by kinetic mixing \citep{millicharge2012}. 
Actually, our analysis further tightens the constraints on millicharged DM \citep{munoz2018insights,berlin2018severely}, as shown in Section~\ref{s3.2} below.

For any BDMS model specified with $m_{\chi}$ and $\sigma_{1}$, we solve the mass threshold of Pop III hosts at any given virialization redshift $z_{\mathrm{vir}}$ with the following steps: We {\it (i)} first derive the pre-virialization thermal and chemical evolution in the select over-dense structure (Sec.~\ref{s2.1}), where the density evolution is modelled with a top-hat model (Sec.~\ref{s2.2}), and then {\it (ii)} use these results to determine the minimum halo mass required to host star formation (Sec.~\ref{s2.3}). The initial conditions for the thermal and chemical evolution are described in Section~\ref{s2.4}.

\begin{figure}
    \vspace{-7pt}
    \hspace{-12pt}
    \includegraphics[width=1.07\columnwidth]{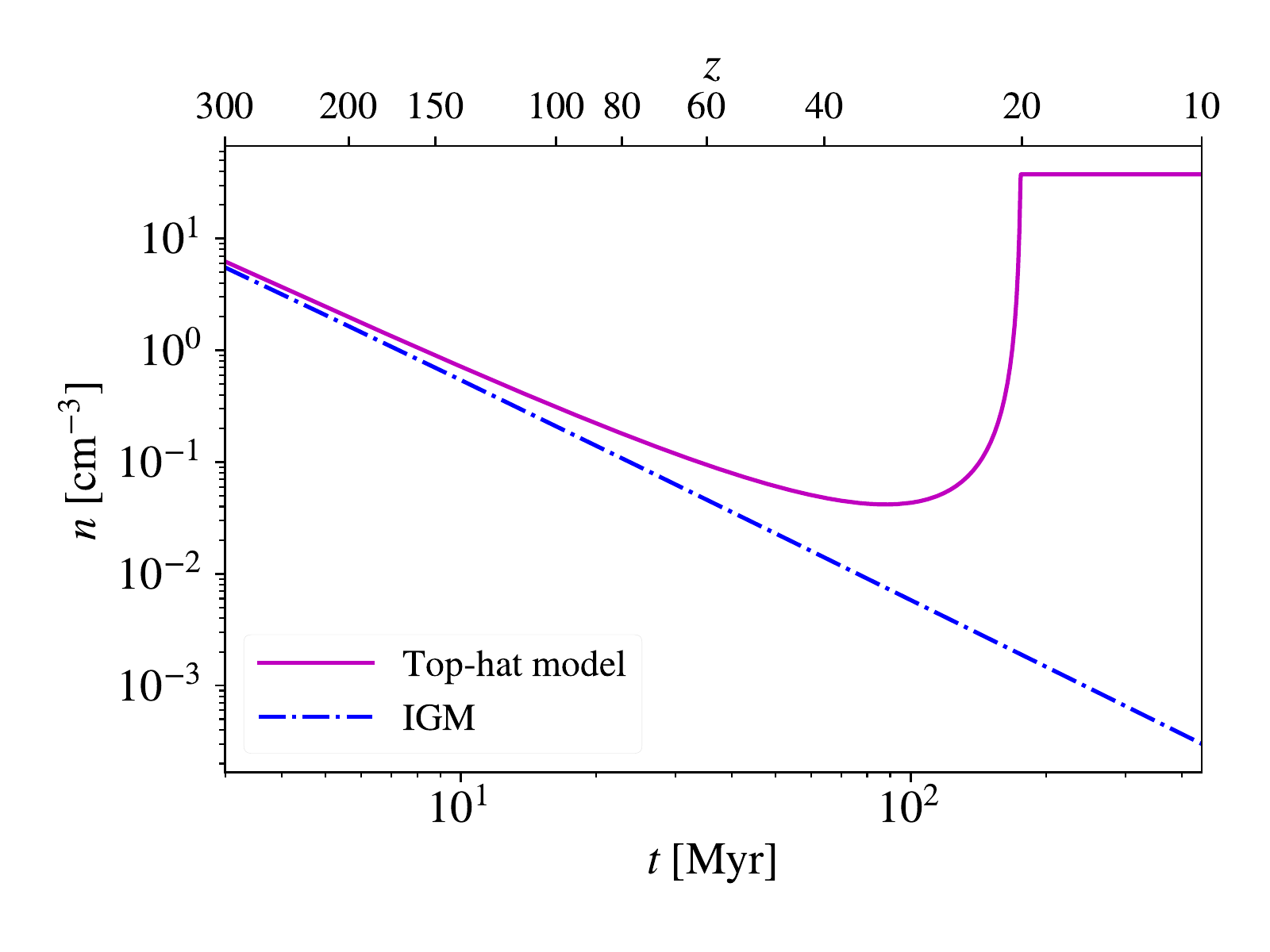}
    \vspace{-25pt}
    \caption{Density evolution for the top-hat model, given by Equations~(\ref{e7})-(\ref{e9}), at $z_{\mathrm{vir}}=20$ (corresponding to $t\simeq 180$~Myr). For comparison, we also show the evolution of the IGM background (dashed-dotted line).}
    \label{f1}
\end{figure}

\subsection{Thermal evolution}
\label{s2.1}
We solve the thermal evolution of primordial gas and DM in over-dense structures with an idealized one-zone model, including the BDMS terms from \citet{BDMS}, while the density evolution is approximated with a top-hat model (see Sec.~\ref{s2.2} for details). The governing equations for the baryon and DM temperatures, $T_{b}$ and $T_{\chi}$, as well as the relative velocity between these two components ($v_{b\chi}$), are
\begin{align}
    \frac{k_{\mathrm{B}}}{\gamma_{b}-1}\frac{dT_{b}}{dt}&= \frac{\Gamma-\Lambda}{n}+\frac{k_{\mathrm{B}}T_{b}}{\rho_{b}}\frac{d\rho_{b}}{dt}+\dot{Q}_{b}\ ,\label{e1}\\
    \frac{k_{\mathrm{B}}}{\gamma_{\chi}-1}\frac{dT_{\chi}}{dt}&= \frac{k_{\mathrm{B}}T_{\chi}}{\rho_{\chi}}\frac{d\rho_{\chi}}{dt}+\dot{Q}_{\chi}\ ,\label{e2}\\
    \frac{dv_{b\chi}}{dt}&=\frac{1}{3}\frac{v_{b\chi}}{\rho_{m}}\frac{d\rho_{m}}{dt}-D\ .\label{e3}
\end{align}
Here $\rho_{i}=(\Omega_{i}/\Omega_{m})\rho_{m}$ denotes the densities of baryons ($i=b$) and DM ($i=\chi$), where $\rho_{m}$ is the total matter density, with $\Omega_{b}=0.048$, $\Omega_{\chi}=0.267$ and $\Omega_{m}=0.315$ in Planck cosmology 
\citep{planck}. Further, $k_{\mathrm{B}}$ is the Boltzmann constant, and we set the adiabatic indices to $\gamma=\gamma_{b}=\gamma_{\chi}=5/3$ for simplicity. 

For baryons ($T_{b}$), the first term on the right-hand side of Equation~(\ref{e1}) represents heating ($\Gamma$) and cooling ($\Lambda$) from baryonic processes, where $n=\rho_{b}/(\mu m_{\mathrm{p}})$ is the number density of gas particles, with $\mu\simeq 1.22$ being the mean molecular weight, and $m_{\mathrm{p}}$ the proton mass ($m_{\mathrm{p}}c^{2}=0.938$~GeV). We only consider the heating by Compton scattering of CMB photons, so that $\Gamma=k_{\mathrm{C}}(T_{\mathrm{CMB}}-T_{b})$ \citep{BDMS}, where $k_{\mathrm{C}}$ is the Compton interaction rate\footnote{$k_{\mathrm{C}}=3.9\times 10^{-20}n_{\mathrm{e}}(1+z)^{4}$ (in c.g.s. units) at redshift $z$, where $n_{\mathrm{e}}$ is the number density of free elections. We have verified that our model can reproduce the thermal IGM evolution in CDM cosmology from \citet{PhysRevD.82.083520}.}, and $T_{\mathrm{CMB}}=T_{\mathrm{CMB},0}(1+z)$ the CMB temperature with a present-day value of $T_{\mathrm{CMB},0}=2.726\ \mathrm{K}$. The cooling function $\Lambda$ is taken from \citet{lithium}, which implements cooling by atoms, ions, free electrons, and the main molecular coolants ($\mathrm{H_{2}}$, $\mathrm{HD}$ and $\mathrm{LiH}$) in primordial gas. The second and third terms on the right hand side of Equation~(\ref{e1}) correspond to the work done to the fluid by adiabatic compression/expansion and energy transfer by BDMS ($\dot{Q}_{b}$). The case of DM ($T_{\chi}$) is similar, without a baryonic heating/cooling term. The time evolution of $v_{b\chi}$ (Equ.~\ref{e3}) is also determined by two terms: one for adiabatic compression/expansion, and the other for BDMS, described with the drag force term $D$. 

Given the masses of baryon and DM particles, $m_{b}$ and $m_{\chi}$, the heat-exchange (energy-transfer) and drag force (momentum-transfer) terms for BDMS take the forms \citep{BDMS}
\begin{align}
    &\dot{Q}_{b}(v_{b\chi};\sigma_{0}, m_{\chi}; m_{b})=\frac{2k_{\mathrm{B}}m_{b}\rho_{\chi}\sigma_{0}(T_{\chi}-T_{b})e^{-r^{2}/2}}{(m_{b}+m_{\chi})^{2} \sqrt{2\pi}u_{\mathrm{th}}^{3}}\notag\\
    &\hspace{60pt} +\frac{m_{b}m_{\chi}\rho_{\chi}}{(m_{b}+m_{\chi})\rho_{m}}v_{b\chi}D(v_{b\chi};\sigma_{0},m_{\chi};m_{b})\ ,\label{e4}\\
    &D(v_{b\chi};\sigma_{0},m_{\chi};m_{b})=\frac{\sigma_{0}\rho_{m}}{(m_{b}+m_{\chi})v_{b\chi}^{2}}F(r)\ ,\label{e5}
\end{align}
under the ideal gas approximation, where $r\equiv v_{b\chi}/u_{\mathrm{th}}$, $u_{\mathrm{th}}\equiv\sqrt{k_{\mathrm{B}}T_{b}/m_{b}+k_{\mathrm{B}}T_{\chi}/m_{\chi}}$ is the typical velocity of relative thermal motion between the two fluids, and
\begin{align}
    F(r)\equiv \text{erf}\left(\frac{r}{\sqrt{2}}\right)-\sqrt{\frac{2}{\pi}}re^{-r^{2}/2}\ .\label{e6}
\end{align}
The heat-exchange term for DM, $\dot{Q}_{\chi}(v_{b\chi};\sigma_{0}, m_{\chi};m_{b})$, can be obtained easily from that for baryons~(Equ.~\ref{e4}) by exchanging the subscripts $b$ and $\chi$. 
To evaluate the relevant terms in Equations~(\ref{e1}) and (\ref{e2}), we only consider BDMS involving hydrogen and helium nuclei, so that $\dot{Q}_{i}\equiv x_{\mathrm{H}}\dot{Q}_{i}(v_{b\chi};\sigma_{0},m_{\chi};m_{\mathrm{p}})+x_{\mathrm{H_{e}}}\dot{Q}_{i}(v_{b\chi};\sigma_{0},m_{\chi};4m_{\mathrm{p}})$ for $i=b,\ \chi$, where $x_{\mathrm{H}}=0.927$ and $x_{\mathrm{H_{e}}}=1-x_{\mathrm{H}}$ are the fractions of hydrogen and helium nuclei in primordial gas. Similarly, for the drag force term in Equation~(\ref{e3}), we have $D\equiv  x_{\mathrm{H}}D(v_{b\chi};\sigma_{0},m_{\chi};m_{\mathrm{p}})+x_{\mathrm{H_{e}}}D(v_{b\chi};\sigma_{0},m_{\chi};4m_{\mathrm{p}})$.

Finally, we need to solve the chemical evolution to evaluate the baryonic heating and cooling terms in Equation~(\ref{e1}). We start with the chemical network in \citet{lithium}, which includes 36 reactions for 17 species \citep{haiman1996,galli1998,bromm2002,mackey2003,johnson2006}. We further include 2 reactions involving CMB photons\footnote{$\mathrm{H^{-}}+h\nu=\mathrm{H}+\mathrm{e^{-}}$ and $\mathrm{H_{2}^{+}}+h\nu=\mathrm{H}+\mathrm{H^{+}}$.} from \citet{galli1998}, which are important for the formation of molecular coolants $\mathrm{H_{2}}$ and $\mathrm{HD}$ at high redshifts ($z\gtrsim 30$). For simplicity, we do not include Lyman-Werner (LW) photons in our network, which can increase the mass threshold of star-forming minihaloes (e.g., \citealt{machacek2001simulations,wise2007suppression,oshea2008}). In general, taking into account LW photons will further tighten the constraints on BDMS. This can be described by the fitting formula $M_{\mathrm{th}}=M_{\mathrm{th},0}\left[1+6.96(4\pi J_{\mathrm{LW},21})^{0.47}\right]$, based on simulation data \citep{machacek2001simulations,Fialkov2014}. Here, $M_{\mathrm{th}}$ and $M_{\mathrm{th},0}$ are the mass thresholds with and without LW fields, $J_{\mathrm{LW},21}\equiv J_{\mathrm{LW}}/(10^{-21}\mathrm{erg\ s^{-1}\ cm^{-2}\ Hz^{-1}\ sr^{-1}})$, and $J_{\mathrm{LW}}$ is the intensity of the LW background. 

\subsection{Top-hat model}
\label{s2.2}
The density evolution in DM haloes is approximated with a generalized top-hat model, which has two parameters: the over-density factor $\Delta$ and virialization redshift $z_{\mathrm{vir}}$. In the standard top-hat model, there is only one free parameter $z_{\mathrm{vir}}$, with a constant $\Delta=\Delta_{V}\simeq 200$ as the typical overdensity of a virialized DM halo (e.g., \citealt{bromm2002}). However, we use a different value of $\Delta$. The reason is that we here focus on the inner core within $\sim 0.1R_{\mathrm{vir}}$, where star formation occurs \citep{bromm2002,druschke18}, with $R_{\mathrm{vir}}$ being the virial radius. As star-forming gas clouds exhibit a quasi-isothermal density profile $\rho\propto r^{-2.2}$ (e.g., \citealt{gao2007first,hirano2015primordial}), the over-density factor for this inner gas core is about 100 times the typical over-density for virialized haloes, $\Delta_{V}=200$, so that we use $\Delta=100\Delta_{V}=2\times 10^{4}$. We follow the treatment in \citet{tegmark1997}, and express the matter density at $z\ge z_{\mathrm{vir}}$ as
\begin{align}
    \rho_{m}(z)=\max\left[\bar{\rho}_{m}(z)(1+\delta),\bar{\rho}_{m}(z_{\mathrm{vir}})\Delta\right]\ ,\label{e7}
\end{align}
where $\bar{\rho}_{m}(z)=2.7\times 10^{-30}(1+z)^{3}\ \mathrm{g\ cm^{-3}}$ is the background average density at redshift $z$, and $\delta\equiv\delta(z,z_{\mathrm{vir}})$ is calculated from \citep{tegmark1997}
\begin{align}
    (1+\delta)&=\frac{9(\alpha-\sin\alpha)^{2}}{2(1-\cos\alpha)^{3}}\ ,\label{e8}\\
    \frac{1+z_{\mathrm{vir}}}{1+z}&=\left(\frac{\alpha-\sin\alpha}{2\pi}\right)^{2/3}\ .\label{e9}
\end{align}
We set $\rho_{m}(z)=\bar{\rho}_{m}(z_{\mathrm{vir}})\Delta$ for $z<z_{\mathrm{vir}}$.
In Fig.~\ref{f1}, we show an example for the top-hat density evolution with $z_{\mathrm{vir}}=20$ (corresponding to $t\simeq 180$~Myr).

\subsection{Mass threshold for star formation}
\label{s2.3}
A minihalo can host Pop~III stars only when the gas is able to efficiently cool, thus triggering run-away collapse. In order to determine the mass threshold $M_{\mathrm{th}}$ above which a halo can form stars, given its virial mass $M$ and virialization redshift $z_{\mathrm{vir}}$, we solve Equations~(\ref{e1})-(\ref{e3}) with the top-hat model from $z_{i}$ to $z_{\mathrm{vir}}$, where $z_{i}=300$ is the initial redshift (see Sec.~\ref{s2.4} for further details). We do not model the virialization process explicitly, but only consider its thermal consequences. That is to say, at $z_{\mathrm{vir}}$, we set both $T_{b}$ and $T_{\chi}$ to the virial temperature\footnote{Here we assume that DM will rapidly thermalize with BDMS during virialization, so that $T_{\chi}\sim T_{b}\sim T_{\mathrm{vir}}$ holds at the end of the virialization process.}
\begin{align}
    T_{\mathrm{vir}}= &\frac{GM\mu m_{\mathrm{p}}}{5k_{\mathrm{B}} R_{\mathrm{vir}}}\simeq 900 \left(\frac{1+z_{\mathrm{vir}}}{21}\right)\left(\frac{M}{10^{6}\ M_{\odot}}\right)^{2/3}\ \mathrm{K}\ ,\label{e10}
\end{align}
for any given $M$. Then we calculate the cooling timescale $t_{\mathrm{cool}}=T_{b}/(dT_{b}/dt)$, where $dT_{b}/dt$ is obtained by evaluating Equation~(\ref{e1}) at $z_{\mathrm{vir}}$. We compare $t_{\mathrm{cool}}$ with the free-fall timescale of the inner core
\begin{align}
    t_{\mathrm{ff}}=\sqrt{\frac{3\pi}{32G\Delta\bar{\rho}_{m}(z_{\mathrm{vir}}) }}\simeq 3 \left(\frac{1+z_{\mathrm{vir}}}{21}\right)^{-3/2}\ \mathrm{Myr}\ ,\label{e11}
\end{align}
and conclude that the halo can undergo star formation when the Rees-Ostriker-Silk (ROS) cooling criterion $t_{\mathrm{cool}}\le t_{\mathrm{ff}}$ \citep{rees1977cooling, silk1977} is met (see also \citealt{Sullivan2018}). In this way, for any given $z_{\mathrm{vir}}$, we can obtain $t_{\mathrm{cool}}/t_{\mathrm{ff}}$ as a function of $M$ by interpolation. Within this framework, $M_{\mathrm{th}}$ is defined by imposing the condition $t_{\mathrm{cool}}/t_{\mathrm{ff}}=1$. 

\begin{figure*}
    \vspace{-7pt}
    \hspace{-13pt}
    \includegraphics[width=2.14\columnwidth]{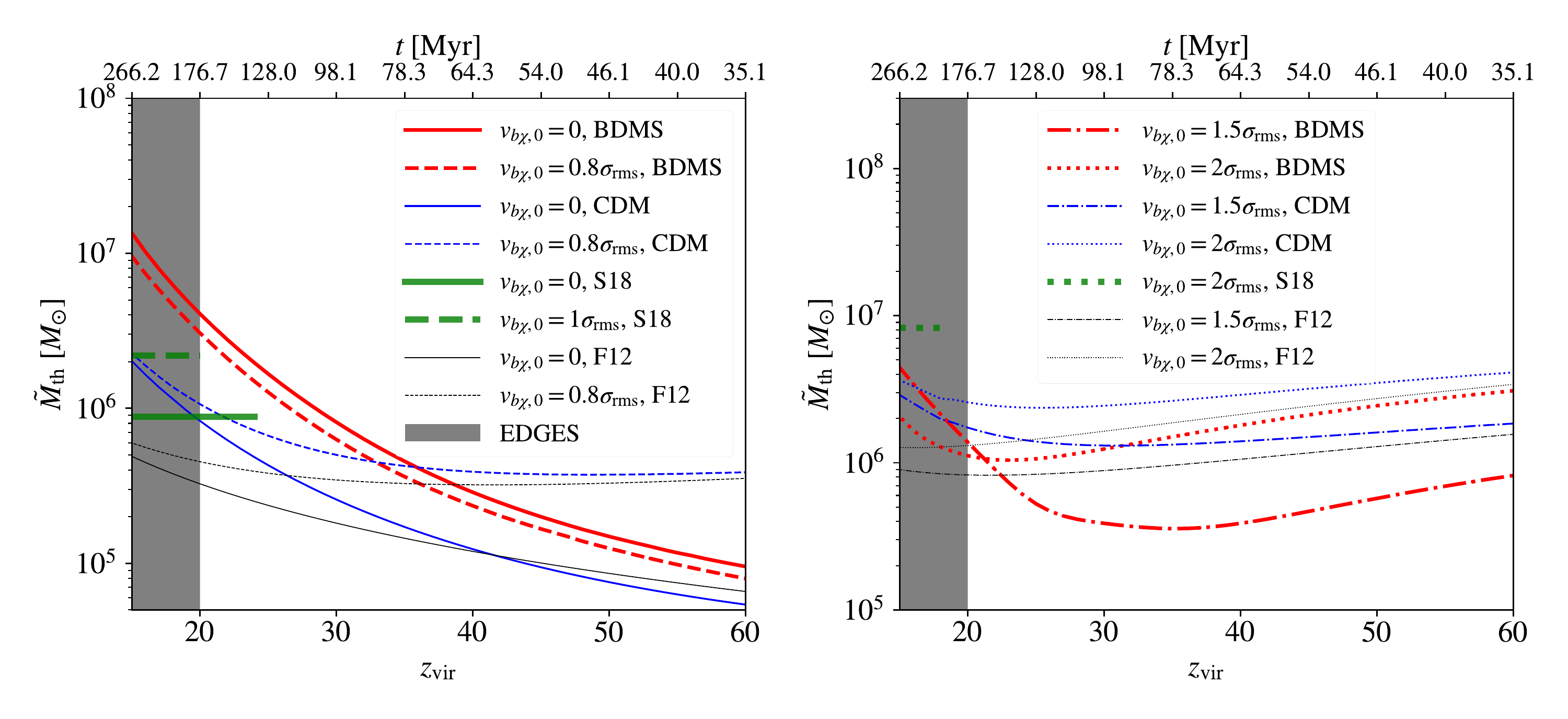}
    \vspace{-15pt}
    \caption{Mass thresholds of Pop~III host haloes as functions of virialization redshift under the fiducial BDMS model (thick lines) with $m_{\chi}c^{2}=0.3$~GeV and $\sigma_{1}=8\times 10^{-20}\ \mathrm{cm^{2}}$, for (left) $v_{b\chi,0}=0$ (solid) and $v_{b\chi,0}=0.8\sigma_{\mathrm{rms}}$ (dashed), as well as (right) $v_{b\chi,0}=1.5\sigma_{\mathrm{rms}}$ (dashed-dotted) and $v_{b\chi,0}=2\sigma_{\mathrm{rms}}$ (dotted), where $\sigma_{\mathrm{rms}}=30\ \mathrm{km\ s^{-1}}$. The CDM counterparts are shown with normal lines. Besides, for the CDM model, we also plot the results based on the best fit (Equ.~\ref{e16}) with $V_{\mathrm{cool},0}=3.714\ \mathrm{km\ s^{-1}}$ and $\alpha=4.015$ from \citealt{fialkov2012} (`F12') in thin curves. The redshift-independent mass thresholds from the simulations of \citealt{Anna2018} (`S18') are shown in ultra thick lines for $v_{b\chi,0}=0$ (solid), $v_{b\chi,0}=1\sigma_{\mathrm{rms}}$ (dashed) and $v_{b\chi,0}=2\sigma_{\mathrm{rms}}$ (dotted). The shaded region denotes the redshift range of the observed 21-cm absorption signal $15\lesssim z\lesssim 20$ from EDGES \citep{nature}.}
    \label{f2}
\end{figure*}

We further make corrections to the threshold masses derived with the above method to take into account the effect of streaming motion between DM and gas. It is found in simulations that with streaming motion, baryon fractions in minihaloes are reduced, and star formation delayed (e.g., \citealt{maio2011impact,greif2011delay,stacy2011effect,naoz12,naoz13}). As a result, the mass threshold for Pop~III host haloes is also increased (e.g., \citealt{fialkov2012,Anna2018}). Although with BDMS, the streaming motion does play a role in our model through Equation~(\ref{e3}) and the dependence on $v_{b\chi}$ of $\dot{Q}_{i}$ ($i=b\ ,\chi$), the effect is not fully captured in the absence of well-modeled dynamics. Therefore, it is necessary to add further corrections. For simplicity, the effect of streaming motion is absorbed into the effective circular velocity 
\begin{align}
    v_{\mathrm{eff}}=\left[v^{2}_{\rm cir}+\beta (\Delta_{V}/\Delta)^{2/3}v_{b\chi,V}^{2}\right]^{1/2}\ ,\label{e13}
\end{align}
where $\beta$ is an adjustable parameter (=0.7 by default), $v_{b\chi,V}\equiv v_{b\chi}(z=z_{\mathrm{vir}})$ is obtained from the one-zone model, $(\Delta_{V}/\Delta)^{2/3}$ a scaling factor to account for the difference in relative velocities for the inner core and the entire halo, and 
\begin{align}
    v_{\mathrm{cir}}&=\sqrt{\frac{GM_{\mathrm{th}}}{R_{\mathrm{vir}}}}\notag\\
    &=5.4\left(\frac{M_{\mathrm{th}}}{10^{6}\ M_{\odot}}\right)^{1/3}\left(\frac{1+z_{\mathrm{vir}}}{21}\right)^{1/2}\ \mathrm{km\ s^{-1}} \label{e14}
\end{align}
is the circular velocity of a halo with $M=M_{\mathrm{th}}$. The corrected threshold mass is then defined as
\begin{align}
    \tilde{M}_{\mathrm{th}}=\left(\frac{v_{\mathrm{eff}}}{v_{\mathrm{cir}}}\right)^{3}M_{\mathrm{th}}\ .\label{e15}
\end{align}
We have verified that in $\Lambda$CDM cosmology, our results with $\beta=0.7$ are consistent with those from the simulations in \citet{Anna2018} (see Fig.~\ref{f2}). Note that, \citet{fialkov2012} has developed a model to describe the dependence of $M_{\mathrm{th}}$ on $v_{bc}$
\begin{align}
    V_{\mathrm{cool}}=\{V^{2}_{\mathrm{cool},0}+[\alpha v_{\mathrm{bc}}]^{2}\}^{1/2}\ ,\label{e16}
\end{align}
where $V_{\mathrm{cool},0}=v_{\mathrm{cir}}(M_{\mathrm{th}})$, $V_{\mathrm{cool}}=v_{\mathrm{cir}}(\tilde{M}_{\mathrm{th}})$, and $v_{\mathrm{bc}}\simeq v_{b\chi,V}/\Delta^{1/3}$. The predictions from this model with the best-fit parameters $V_{\mathrm{cool},0}=3.714\ \mathrm{km\ s^{-1}}$ and $\alpha=4.015$ are consistent with our results at $z\gtrsim 40$, as shown in Fig.~\ref{f2}. Actually, it can be easily shown from Equations~(\ref{e13})-(\ref{e16}) that the choice of $\alpha=\beta^{1/2}\Delta_{V}^{1/3}$ and $\beta=0.7$ corresponds to $\alpha=4.89$ for $\Delta_{V}=200$, within the range $\alpha\sim 4-6$ calibrated with simulations \citep{Fialkov2014}. 

Finally, we impose an upper bound to $\tilde{M}_{\mathrm{th}}$ as $M_{\max}=10 M_{2}$, where (e.g., \citealt{yoshida2003simulations,trenti2009formation})
\begin{align}
M_{2}\equiv M_{2}(z_{\mathrm{vir}})&\simeq 8.2\times 10^{6}\left(\frac{1+z_{\mathrm{vir}}}{21}\right)^{-3/2}\ M_{\odot}\label{e12}
\end{align}
is the mass threshold for atomic cooling haloes \citep{OhHaiman2002}. Here we assume that haloes with $M>M_{\max}$ can cool efficiently anyway by atomic cooling, which is activated by structure formation shocks and not captured by our simple treatment of the virialization process without dynamical ionization. 

\subsection{Initial conditions}
\label{s2.4}
Our calculations of the thermal, chemical, and density evolution of over-dense structures start at an initial redshift $z_{i}=300$. The choice of $z_{i}$ is based on the fact that for the parameter space of greatest concern here, significant differences between the standard CDM and BDMS models only occur at $z\ll z_{i}$ (see below)\footnote{Our chemical network is incomplete and cannot reproduce the chemical evolution in CDM cosmology at high redshifts ($z\gg z_{i}$), but the results are consistent with those in \citet{galli2013dawn} for $z<z_{i}$.}. 
The chemical network is initialized with the abundance values from \citet{galli2013dawn} at $z_{i}=300$, as summarized in Table~\ref{t1}.

\begin{table}
    \centering
    \caption{Initial abundances of select primordial chemical species at $z_{i}=300$, taken from \citet{galli2013dawn}.}
    \begin{tabular}{cccccc}
         \hline
        $\mathrm{[H^{+}/H]}$ & $\mathrm{[H^{-}/H]}$ & $\mathrm{[H_{2}/H]}$ & $\mathrm{[H_{2}^{+}/H]}$ & $\mathrm{[D/H]}$ \\
        \hline
        $5\times 10^{-4}$ & $10^{-18}$ & $10^{-11}$ & $5\times 10^{-16}$ & $4\times 10^{-5}$ \\
        \hline
    \end{tabular}
    \label{t1}
\end{table}

The initial conditions for $T_{b}$, $T_{\chi}$ and $v_{b\chi}$ at $z_{i}$ are generated with Equations~(\ref{e1})-(\ref{e3}) for the IGM background, starting the calculation at recombination, $z_{\mathrm{rec}}=1100$, now turning off the chemical network, and assuming that BDMS effects are not significant at such high redshifts. Furthermore, instead of using the top-hat model, we simply set $\rho_{m}=\bar{\rho}_{m}$, as the density perturbation has not grown significantly to deviate from the background. 

At $z_{\mathrm{rec}}=1100$, the magnitude of the streaming motion between DM and gas $v_{b\chi,0}$ is an adjustable parameter. Such streaming motion can be coherent over large (Mpc) scales, and $v_{b\chi,0}$ follows a multi-variate Gaussian distribution (e.g., \citealt{PhysRevD.82.083520,Fialkov2014}). The 3-D standard deviation of this distribution is $\sigma_{\mathrm{rms}}=30\ \mathrm{km\ s^{-1}}$, and $0.8\sigma_{\mathrm{rms}}$ serves as a typical value of $v_{b\chi,0}$, around which the contribution to the overall halo mass function is largest \citep{Anna2019}. 

For $T_{b}$ in the IGM background, we use the fitting formula from \citet{PhysRevD.82.083520}
\begin{align}
    T_{b}(a)=\frac{T_{\mathrm{CMB},0}}{a}\left[1+\frac{a/a_{1}}{1+(a_{2}/a)^{3/2}}\right]^{-1}\ ,\label{e17}
\end{align}
where $a=1/(1+z)$ is the scale factor, $a_{1}=1/119$, and $a_{2}=1/115$. While $T_{\chi}$ is estimated by
\begin{align}
    T_{\chi}(z)=
    \begin{cases}
    &T_{\mathrm{CMB},0}(1+z),\hspace{48pt} z\ge z_{\mathrm{crit}}\ , \\
    &T_{\chi}(z_{\mathrm{crit}})\left(\frac{1+z}{1+z_{\mathrm{crit}}}\right)^{2},\hspace{30pt} z<z_{\mathrm{crit}}\ ,
    \end{cases}\label{e18}
\end{align}
where $z_{\mathrm{crit}}=2m_{\chi}c^{2}/(3k_{\mathrm{B}}T_{\mathrm{CMB},0})-1$ is the critical redshift below which DM particles become non-relativistic. For instance, $T_{b}=2991$~K and $T_{\chi}=3.9\times 10^{-6}$~K at $z_{\mathrm{rec}}=1100$, given $m_{\chi}c^{2}=0.3$~GeV.

\section{Results}
\label{s3}

\subsection{The fiducial BDMS model}
\label{s3.1}

\begin{figure}
    \vspace{-7pt}
    \hspace{-12pt}
    \includegraphics[width=1.07\columnwidth]{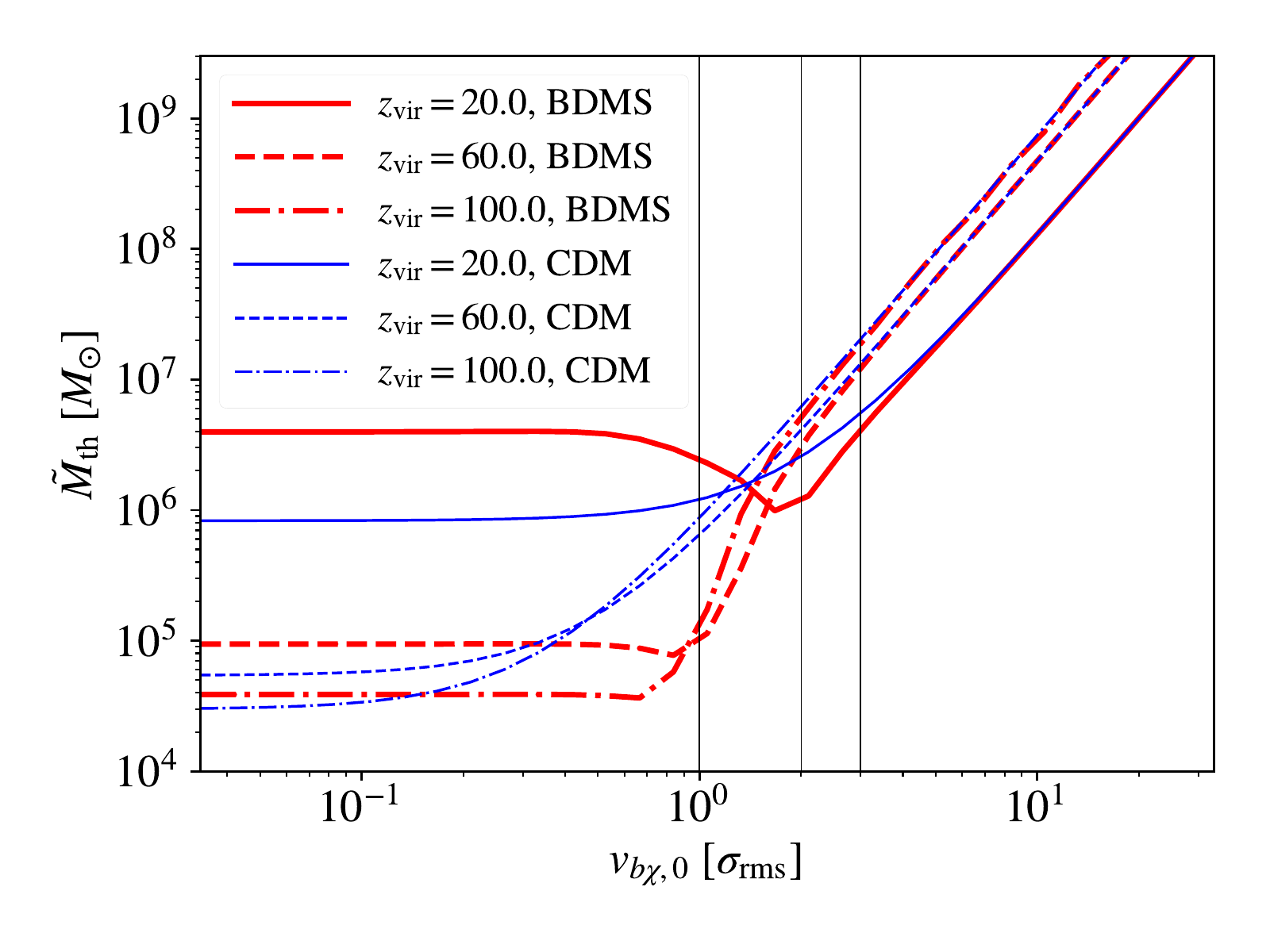}
    \vspace{-25pt}
    \caption{Mass thresholds of Pop~III host haloes as functions of initial streaming motion velocity for $z_{\mathrm{vir}}=20$ (solid), $z_{\mathrm{vir}}=60$ (dashed) and $z_{\mathrm{vir}}=100$ (dotted). The results for the fiducial BDMS model with $m_{\chi}c^{2}=0.3$~GeV and $\sigma_{1}=8\times 10^{-20}\ \mathrm{cm^{2}}$ are shown with thick curves, while those for the CDM model with normal curves. The thin vertical lines show the locations of 1, 2 and 3 $\sigma_{\mathrm{rms}}$, where $\sigma_{\mathrm{rms}}=30\ \mathrm{km\ s^{-1}}$.}
    \label{f3}
\end{figure}

\begin{figure}
    \vspace{-7pt}
    \hspace{-12pt}
    \includegraphics[width=1.07\columnwidth]{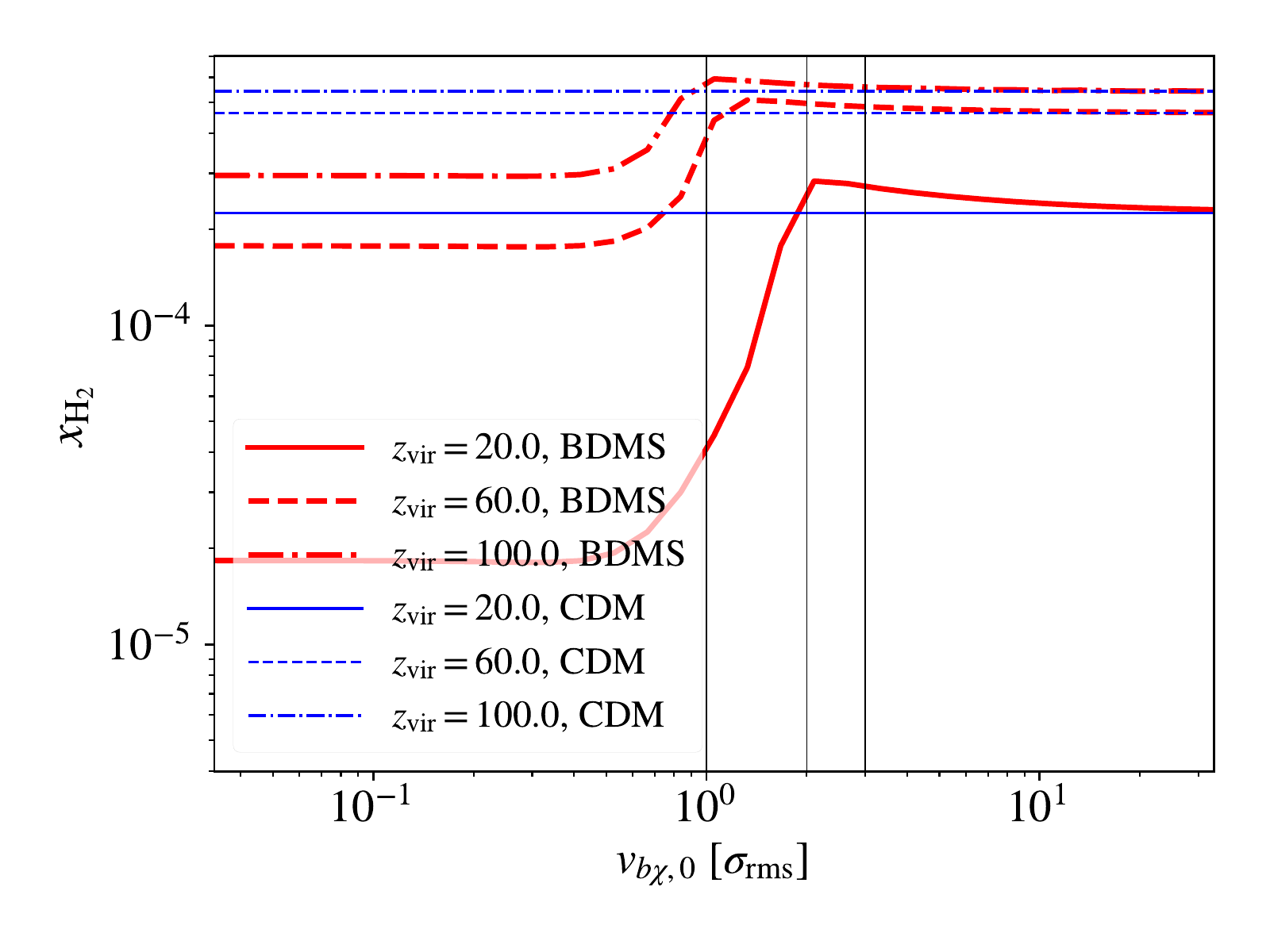}
    \vspace{-25pt}
    \caption{Molecular hydrogen abundances at $z_{\mathrm{vir}}$ as functions of initial streaming motion velocity for $z_{\mathrm{vir}}=20$ (solid), $z_{\mathrm{vir}}=60$ (dashed) and $z_{\mathrm{vir}}=100$ (dotted). Again, the results for the fiducial BDMS model with $m_{\chi}c^{2}=0.3$~GeV and $\sigma_{1}=8\times 10^{-20}\ \mathrm{cm^{2}}$ are shown with thick curves, while those for the CDM model with normal curves. The thin vertical lines show the locations of 1, 2 and 3 $\sigma_{\mathrm{rms}}$, where $\sigma_{\mathrm{rms}}=30\ \mathrm{km\ s^{-1}}$.}
    \label{f4}
\end{figure}

We first explore the fiducial model of BDMS with $m_{\chi}c^{2}=0.3$~GeV and $\sigma_{1}=8\times 10^{-20}\ \mathrm{cm^{2}}$. This model predicts a 21-cm absorption peak with $\langle\delta T\rangle=-500$~mK at $z=17$ \citep{nature1}, which matches the most likely observed value from EDGES \citep{nature}. Fig.~\ref{f2} shows the mass thresholds of Pop~III host haloes under this model as functions of virialization redshift (thick lines), in comparison with those in the standard $\Lambda$CDM model (normal lines), for different magnitudes of initial streaming motion velocity. We also compare our results for the CDM model with the simulation results (ultra-thick lines) from \citet{Anna2018}, as well as the predictions of the best-fit model (Equ.~\ref{e16}) with $V_{\mathrm{cool},0}=3.714\ \mathrm{km\ s^{-1}}$ and $\alpha=4.015$ (thin lines) in \citet{fialkov2012} and find that the differences are within a factor of 3. Without streaming motion, the mass threshold is always enhanced by BDMS, and the enhancement increases as redshift decreases, reaching a factor of 5 at the onset of the 21-cm absorption signal $z_{\mathrm{vir}}\sim 20$. However, with moderate levels of streaming motion ($0.5\sigma_{\mathrm{rms}}\lesssim v_{b\chi,0}\lesssim 2\sigma_{\mathrm{rms}}$), the mass threshold can be reduced due to BDMS at high redshifts $z_{\mathrm{vir}}>z_{\mathrm{th}}$, where the threshold redshift $z_{\mathrm{th}}$ decreases with $v_{b\chi,0}$. Actually, for $v_{b\chi,0}\gtrsim 1.5\sigma_{\mathrm{rms}}$, the mass threshold with BDMS is always lower than that in the CDM model at $z_{\mathrm{vir}}\gtrsim 20$. In the redshift range of the observed 21-cm absorption signal, $15\lesssim z\lesssim 20$ \citep{nature}, the mass threshold is enhanced by a factor of a few, under the representative condition $v_{v\chi,0}=0.8\sigma_{\mathrm{rms}}$. 

Fig.~\ref{f3} shows the mass thresholds as functions of initial streaming motion velocity, at different redshifts. In general, for a given $z_{\mathrm{vir}}$, as $v_{b\chi,0}$ increases, the mass threshold with BDMS present starts above the CDM comparison value, then drops below, and finally becomes equal to it. This trend results from an inverse trend in molecular hydrogen abundances, as shown in Fig.~\ref{f4}. The reason is that the mass threshold tends to decrease with higher $\mathrm{H_{2}}$ abundances, as $\mathrm{H_{2}}$ is the main coolant in minihaloes, and more $\mathrm{H_{2}}$ leads to more efficient cooling under the same conditions. Another factor that shapes the trend in Fig.~\ref{f3} is that BDMS produces friction between the two fluids and facilitates the decay of $v_{b\chi}$. Therefore, under mild initial streaming motions (e.g., $1\sigma_{\mathrm{rms}}\lesssim v_{b\chi,0}\lesssim 3\sigma_{\mathrm{rms}}$ for $z_{\mathrm{vir}}=20$), the effect of streaming motion is reduced in BDMS models, which can lead to lower mass thresholds compared with the CDM model, where $v_{b\chi}$ can only decay with adiabatic expansion.

\begin{figure}
    \vspace{-5pt}
    \hspace{-12pt}
    \includegraphics[width=1.06\columnwidth]{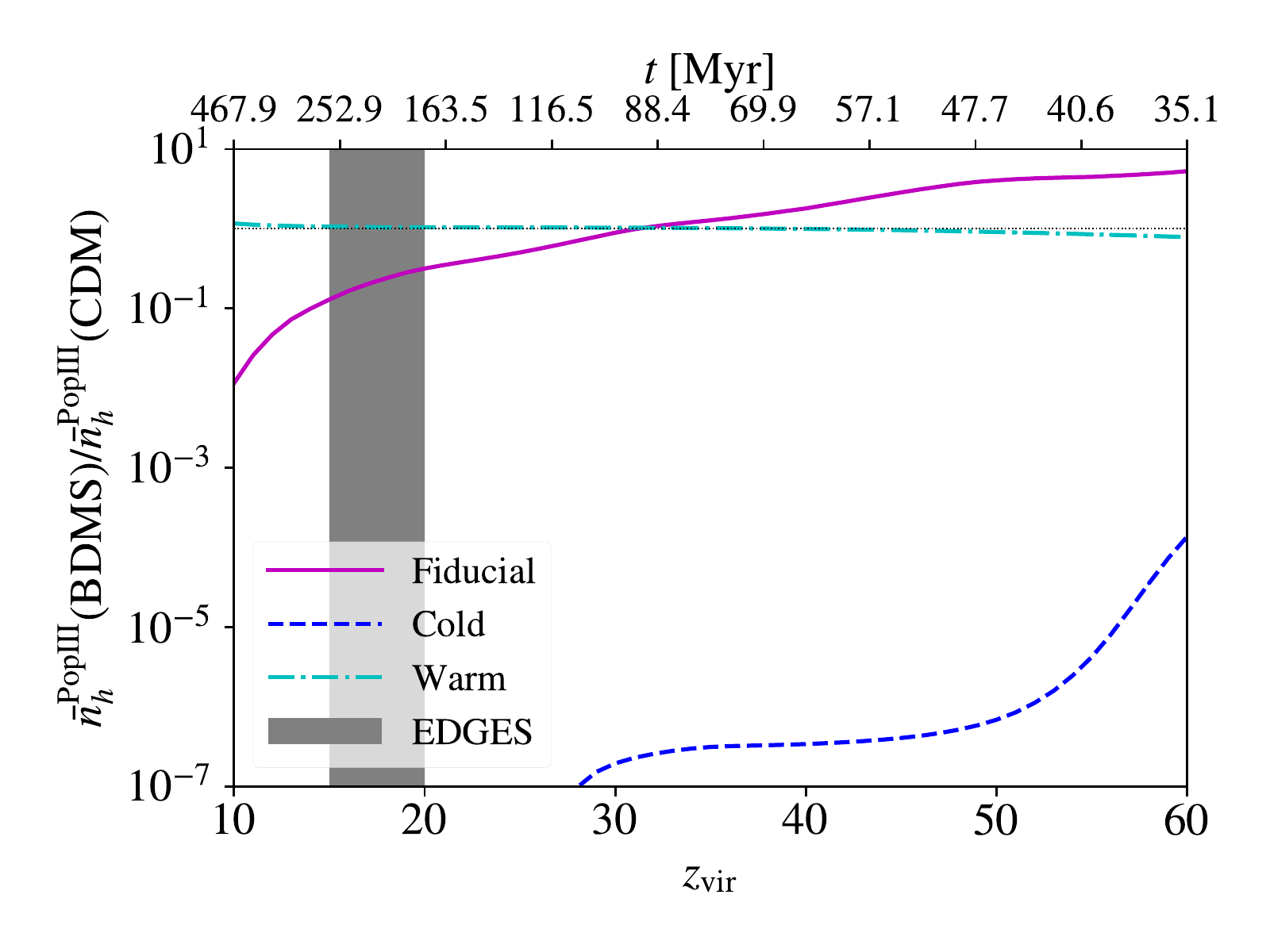}
    \vspace{-20pt}
    \caption{Evolution of the ratio $\bar{n}_{h}^{\mathrm{Pop III}}(\mathrm{BDMS})/\bar{n}_{h}^{\mathrm{Pop III}}(\mathrm{CDM})$ with $z_{\mathrm{vir}}$, for the fiducial BDMS model with $m_{\chi}c^{2}=0.3$~GeV and $\sigma_{1}=8\times 10^{-20}\ \mathrm{cm^{2}}$ (solid), the typical `cold' model with $m_{\chi}c^{2}=10^{-3}$~GeV and $\sigma_{1}=1\times 10^{-18}\ \mathrm{cm^{2}}$ (dashed) and the typical `warm' model with $m_{\chi}c^{2}=10$~GeV and $\sigma_{1}=1\times 10^{-20}\ \mathrm{cm^{2}}$ (dashed-dotted).}
    \label{Nhrat}
\end{figure}

Based on the above results for $\tilde{M}_{\mathrm{th}}(v_{b\chi,0},z_{\mathrm{vir}})$ (Fig.~\ref{f3}), we estimate how the abundance of Pop~III host haloes is affected by BDMS. Assuming that the effect of BDMS on the statistics of DM structures is negligible, we use the \textsc{python} package \href{http://hmf.readthedocs.io/en/latest/index.html}{\texttt{hmf}} \citep{murray2013hmfcalc} to calculate the halo mass functions, with the default fitting model from \citet{tinker2008toward}, and treat the BDMS models as WDM models with thermal WDM masses of the corresponding $m_{\chi}$ values \citep{bode2001halo,viel2005constraining}. Given $z_{\mathrm{vir}}$, we first obtain the number density of Pop~III host haloes, $n^{\mathrm{Pop III}}_{h}$, by integrating the halo mass function in the mass range $[\tilde{M}_{\mathrm{th}}(v_{b\chi,0},z_{\mathrm{vir}}),\ M_{2}(z_{\mathrm{vir}})]$ on a grid of $v_{b\chi}$ covering the interval $[0,5\sigma_{\mathrm{rms}}]$, and then derive the cosmic average $\bar{n}_{h}^{\mathrm{Pop III}}$ from the probability distribution of $v_{b\chi,0}$, which is a multivariate
Gaussian $\mathcal{P}(v_{b\chi,0})\propto v_{b\chi,0}^{2}\exp[-3v_{b\chi,0}^{2}/(2\sigma_{\mathrm{rms}}^{2})]$. The resulting number density ratio $f_{n_{h}}=\bar{n}_{h}^{\mathrm{Pop III}}(\mathrm{BDMS})/\bar{n}_{h}^{\mathrm{Pop III}}(\mathrm{CDM})$ as a function of $z_{\mathrm{vir}}$ is shown in Fig.~\ref{Nhrat}, for the fiducial BDMS model. It turns out that $f_{n_{h}}\lesssim 1$ for $z_{\mathrm{vir}}\lesssim 30$, and $f_{n_{h}}\sim 0.1-0.3$ in the epoch of the EDGES signal, $z_{\mathrm{vir}}\sim 15- 20$. Since sufficient Pop~III star formation is necessary to produce the early 21-cm Wouthuysen-Field coupling inferred by EDGES \citep{Hirano2018,Anna2019}, the suppression of Pop~III halo abundances in the fiducial BDMS model indicates that a higher star-formation efficiency, by a factor of 3 to 10 compared to the CDM case, is required to compensate. 

For comparison, we also plot the results of a typical `cold' model with $m_{\chi}c^{2}=10^{-3}$~GeV and $\sigma_{1}=1\times 10^{-18}\ \mathrm{cm^{2}}$, and a typical `warm' model with $m_{\chi}c^{2}=10$~GeV and $\sigma_{1}=1\times 10^{-20}\ \mathrm{cm^{2}}$ in Fig.~\ref{Nhrat}. In the `cold' model, the number density of Pop~III host haloes is reduced by at least 5 orders of magnitude for $z_{\mathrm{vir}}\lesssim 45$, implying that almost no star formation can happen. While in the `warm' model, $f_{n_{h}}$ is always close to unity, slightly exceeding it for $z_{\mathrm{vir}}\lesssim 35$, and the number density of Pop~III hosts is increased by about $5\%$ at $z_{\mathrm{vir}}\sim 15- 20$. However, this `warm' model cannot produce a strong 21-cm absorption signal, as shown below.

 \begin{figure}
    \vspace{-7pt}
    \hspace{-13pt}
    \includegraphics[width=1.07\columnwidth]{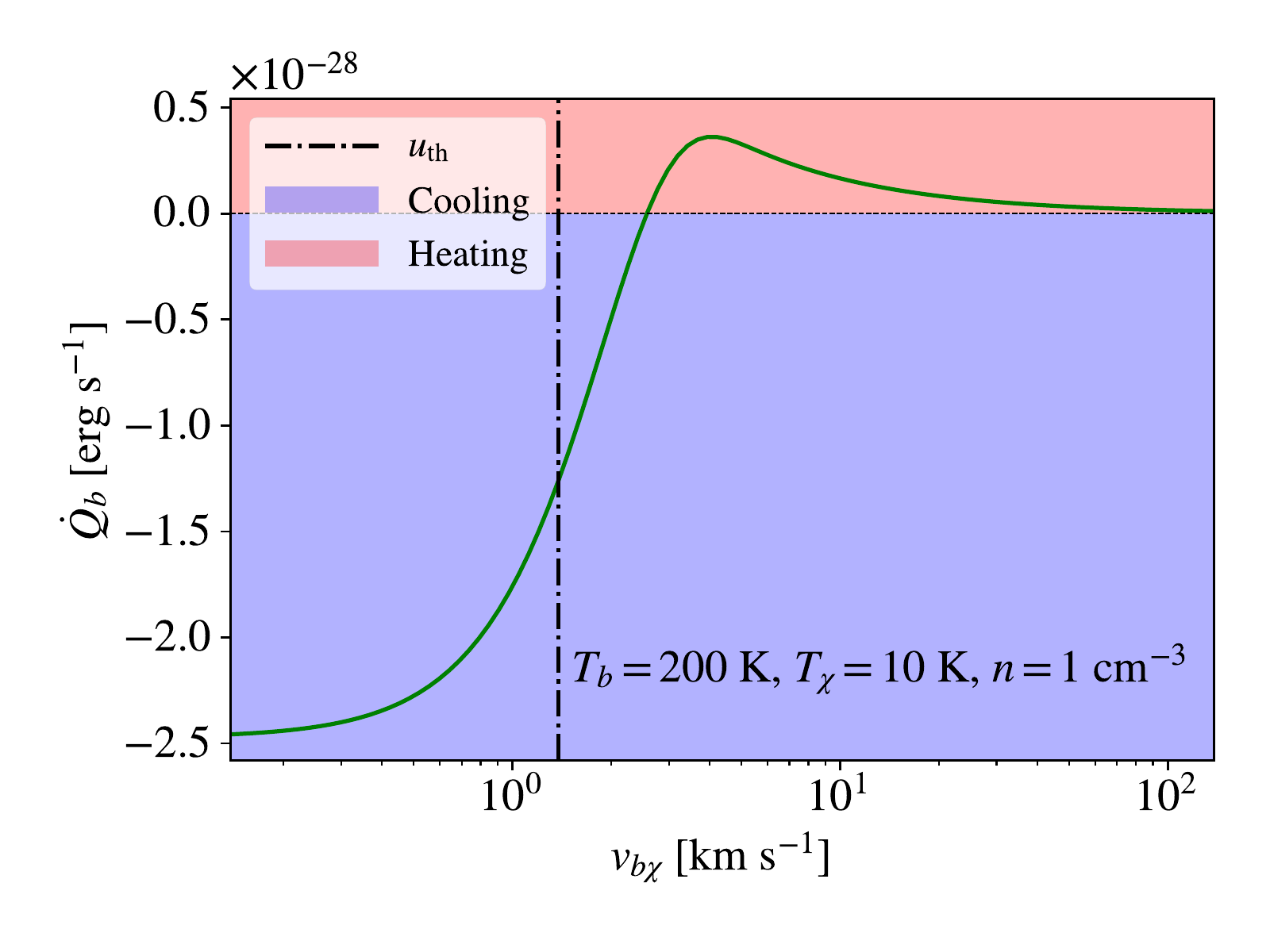}
    \vspace{-20pt}
    \caption{BDMS energy transfer rate per hydrogen nuclei as a function of the relative velocity (solid line), with $T_{b}=200$~K, $T_{\chi}=10$~K, and $n=1\ \mathrm{cm^{-3}}$, under the fiducial BDMS model with $m_{\chi}c^{2}=0.3$~GeV and $\sigma_{1}=8\times 10^{-20}\ \mathrm{cm^{2}}$. The typical velocity of the thermal relative motion between the two fluids $u_{\mathrm{th}}$ is shown with the dashed-dotted vertical line. The upper region with $\dot{Q}_{b}>0$ corresponds to heating of primordial gas by BDMS, while the lower region with $\dot{Q}_{b}<0$ to cooling.}
    \label{f5}
\end{figure}

\begin{figure*}
\subfloat[$v_{b\chi,0}=0$]{\hspace{-11pt}\includegraphics[width= 1.09\columnwidth]{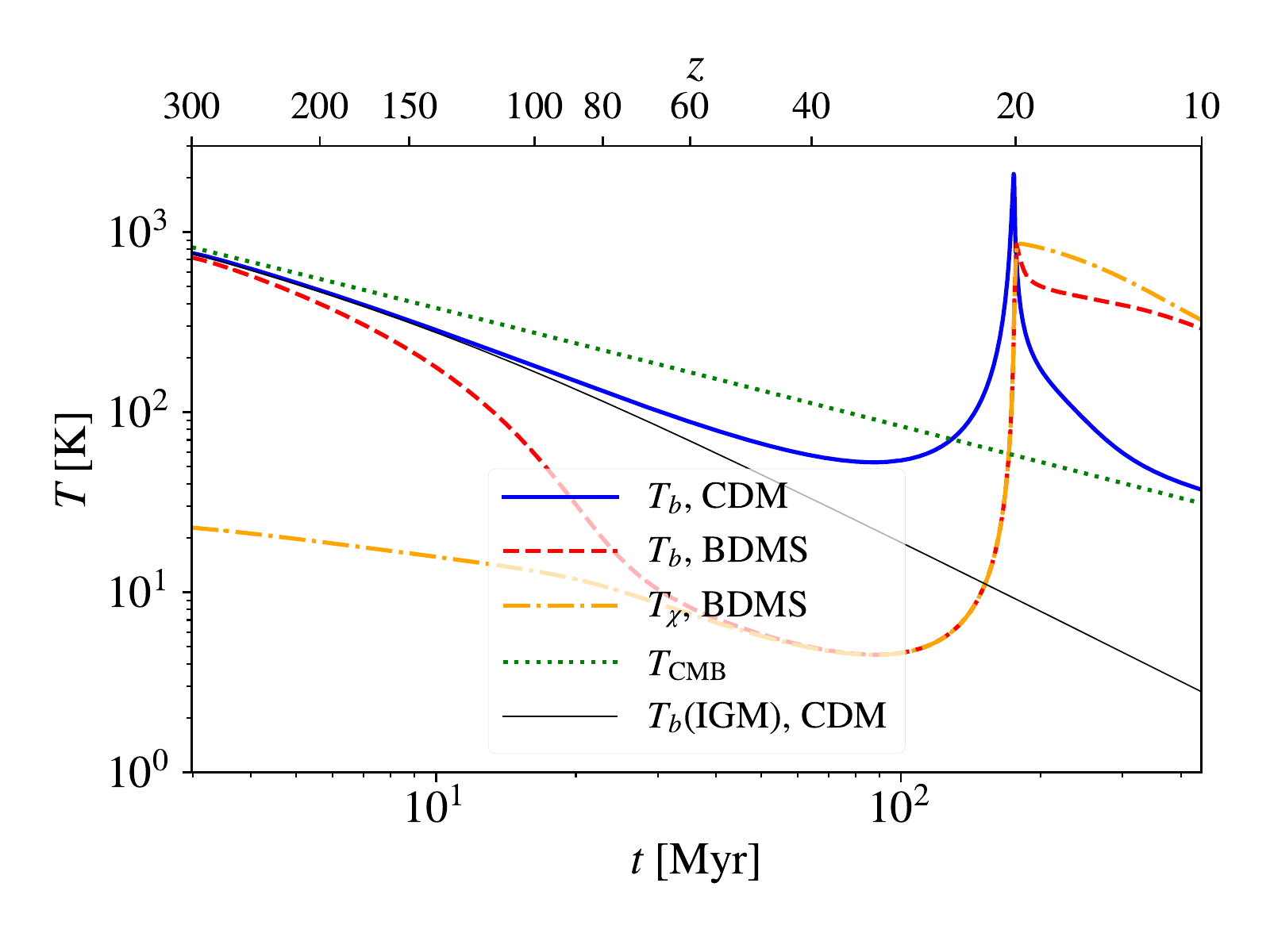}}
\subfloat[$v_{b\chi,0}=0.8\sigma_{\mathrm{rms}}$]{\includegraphics[width= 1.09\columnwidth]{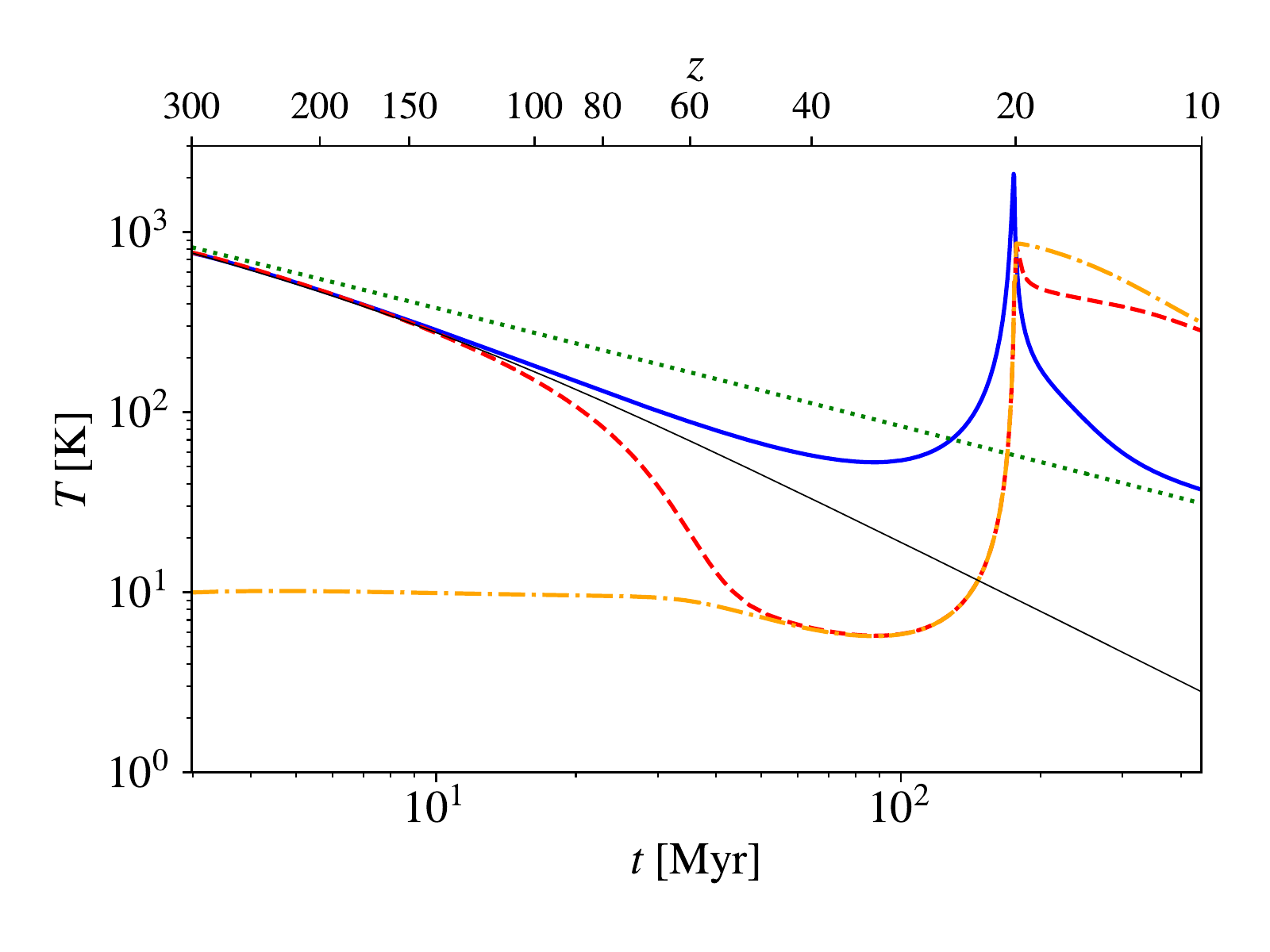}}
\vspace{-15pt}
\subfloat[$v_{b\chi,0}=1.5\sigma_{\mathrm{rms}}$]{\hspace{-11pt}\includegraphics[width= 1.09\columnwidth]{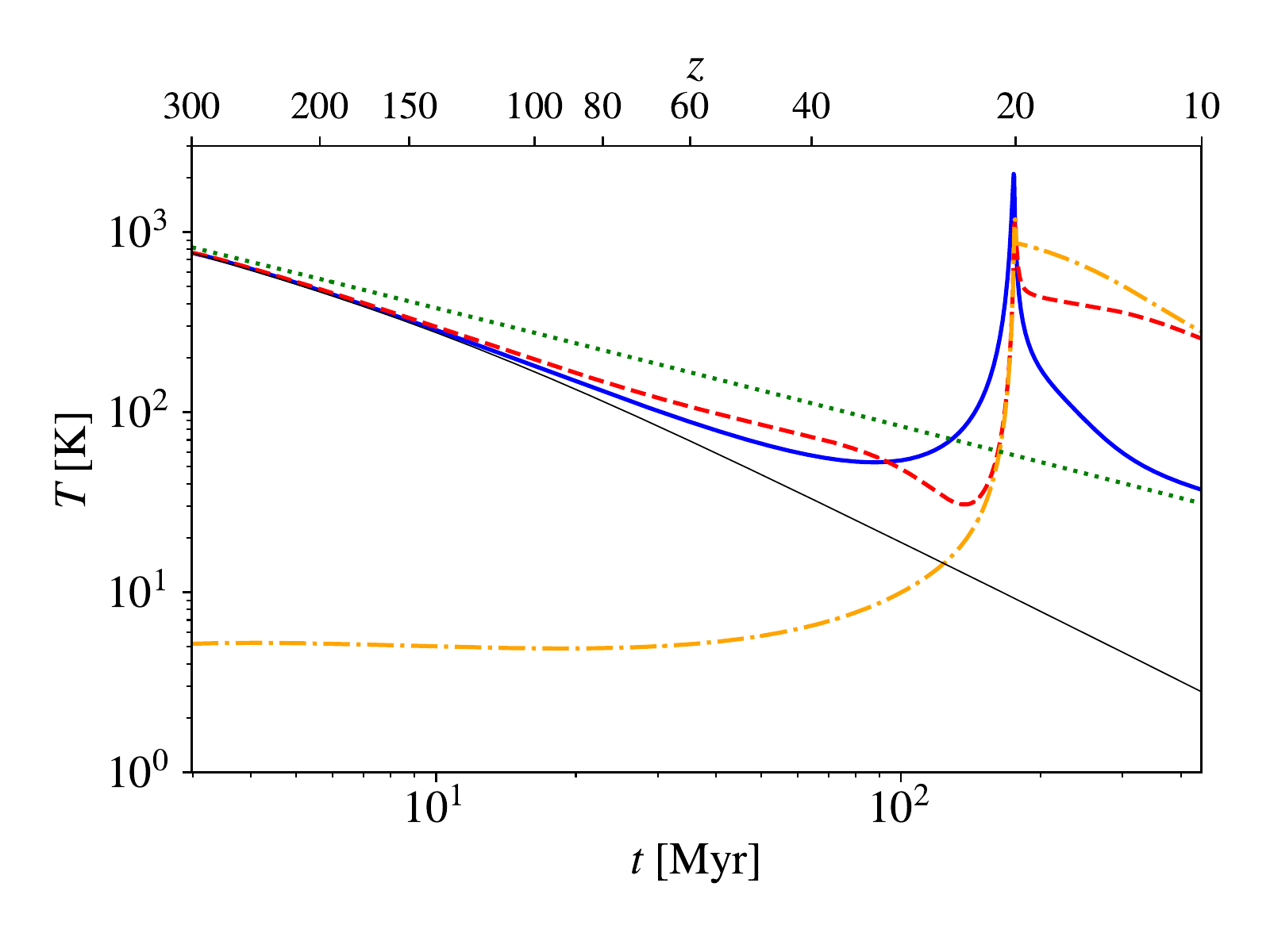}}
\subfloat[$v_{b\chi,0}=2\sigma_{\mathrm{rms}}$]{\includegraphics[width= 1.09\columnwidth]{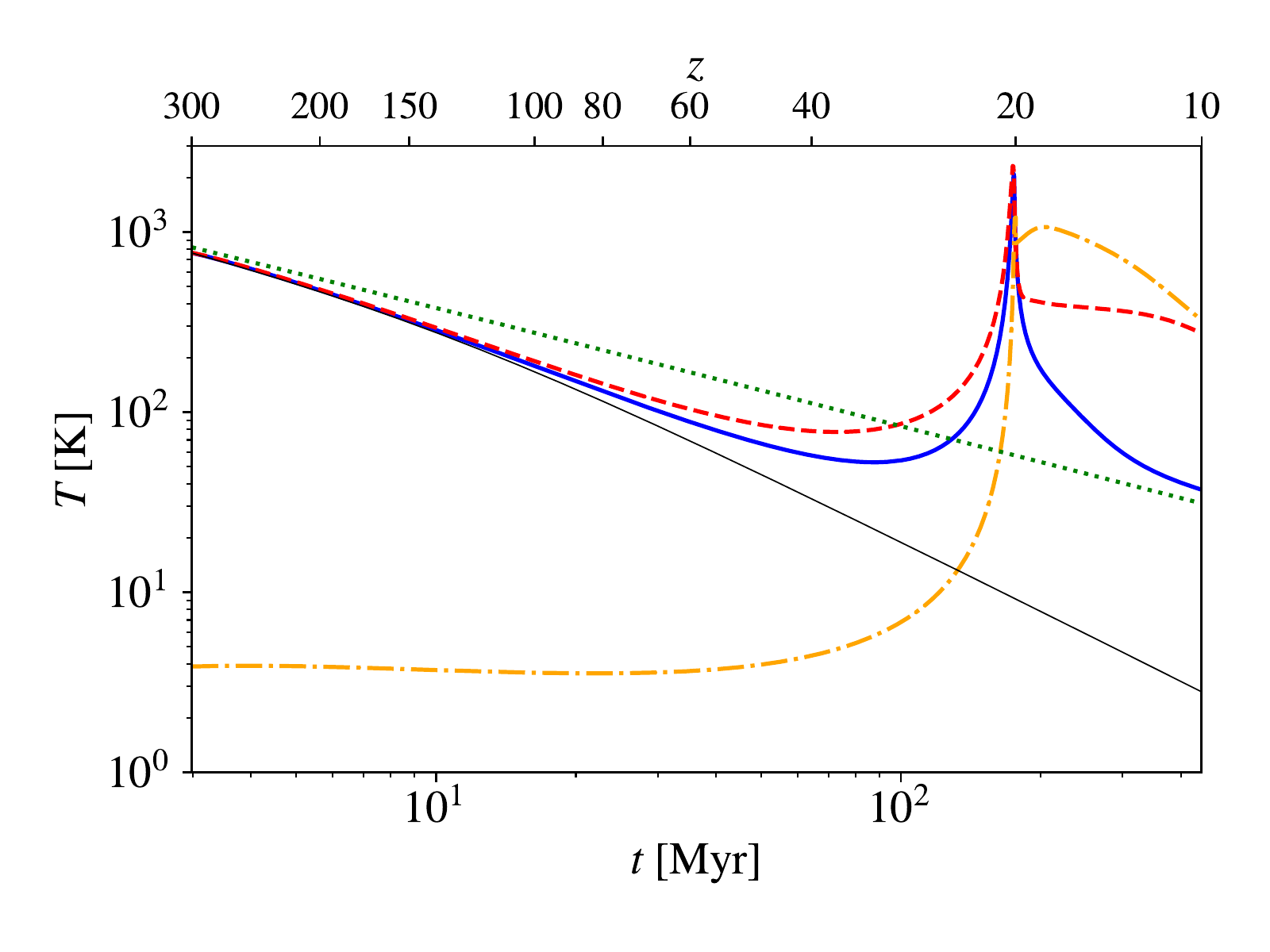}}
\vspace{-15pt}
\caption{Thermal histories for a halo with $M=10^{6}\ M_{\odot}$ and $z_{\mathrm{vir}}=20$ (corresponding to $t\simeq 180$~Myr) under the fiducial BDMS model with $m_{\chi}c^{2}=0.3$~GeV and $\sigma_{1}=8\times 10^{-20}\ \mathrm{cm^{2}}$ (dashed for $T_{b}$ and dashed-dotted for $T_{\chi}$), for (a) $v_{b\chi,0}=0$ (b) $v_{b\chi,0}=0.8\sigma_{\mathrm{rms}}$, (c) $v_{b\chi,0}=1.5\sigma_{\mathrm{rms}}$, and (d) $v_{b\chi,0}=2\sigma_{\mathrm{rms}}$, where $\sigma_{\mathrm{rms}}=30\ \mathrm{km\ s^{-1}}$. The evolution of $T_{b}$ in CDM cosmology is shown with the solid curves, while the CMB temperature $T_{\mathrm{CMB}}$ with the dotted lines. For comparison, we also plot the thermal evolution of the IGM within CDM cosmology, given by the fitting formula (Equ.~\ref{e17}) from \citet{PhysRevD.82.083520} with the thin solid curve.}
\label{f6}
\end{figure*}

The relation between the molecular hydrogen abundance and the initial streaming velocity $v_{b\chi,0}$ shown in Fig.~\ref{f4} can be further understood as follows. In Fig.~\ref{f5}, we illustrate the BDMS energy transfer rate for baryons, $\dot{Q}_{b}$, as a function of the relative velocity $v_{b\chi}$, with $T_{b}=200$~K, $T_{\chi}=10$~K and $n=1\ \mathrm{cm^{-3}}$. It turns out that $\dot{Q}_{b}<0$ for $v_{b\chi}\lesssim u_{\mathrm{th}}$, $\dot{Q}>0$ for $v_{b\chi}\gtrsim u_{\mathrm{th}}$, and $\dot{Q}\simeq 0$ for $v_{b\chi}\gtrsim 10 u_{\mathrm{th}}$. Generally speaking, if the initial streaming velocity is very high, the BDMS effect is negligible. This is why the mass thresholds and $\mathrm{H_{2}}$ abundances converge to the CDM values for $v_{b,\chi,0}\gtrsim 3\sigma_{\mathrm{rms}}$ (see Fig.~\ref{f3} and \ref{f4}). When $v_{b\chi,0}$ is around $0.3-3\sigma_{\mathrm{rms}}$, early on BDMS introduces a heating term for baryons (by friction), and it will subsequently become a cooling term, when $v_{b\chi}$ has decayed significantly due to the drag force. And finally, the two components tend to reach thermal equilibrium at $T_{\mathrm{fin}}\simeq T_{b}(\mathrm{CDM})/[1+(m_{\chi}c^{2}/6\ \mathrm{GeV})^{-1}]$ \citep{nature1}. The competition between heating and cooling in the thermal history affects the chemical evolution, which can result in enhanced or reduced $\mathrm{H}_{2}$ abundances. In this regime, the higher the $v_{b\chi,0}$, the longer it will take for $v_{b\chi}$ to decay, and the prolonged heating will facilitate $\mathrm{H_{2}}$ formation, thus reducing the mass threshold. For instance, Fig.~\ref{f6} shows the thermal histories for a halo with $M=10^{6}\ M_{\odot}$ and $z_{\mathrm{vir}}=20$ (corresponding to $t\simeq 180$~Myr) within the fiducial BDMS model (dashed and dashed-dotted), in comparison with the CDM model (solid), for (a) $v_{b\chi,0}=0$ (b) $v_{b\chi,0}=0.8\sigma_{\mathrm{rms}}$, (c) $v_{b\chi,0}=1.5\sigma_{\mathrm{rms}}$, and (d) $v_{b\chi,0}=2\sigma_{\mathrm{rms}}$. It turns out that cooling dominates before virialization for $v_{b\chi,0}\le 0.8\sigma_{\mathrm{rms}}$. While for $v_{b\chi,0}=1.5\sigma_{\mathrm{rms}}$, the transition from heating to cooling happens at $t\sim 70$~Myr. When $v_{b\chi,0}=2\sigma_{\mathrm{rms}}$, BDMS always heats up the gas for $z>z_{\mathrm{vir}}$ (i.e., $t\lesssim 180$~Myr), since $v_{b\chi}> u_{\mathrm{th}}$ always holds. 

\begin{figure*}
\subfloat[$v_{b\chi,0}=0$]{\hspace{-11pt}\includegraphics[width= 1.09\columnwidth]{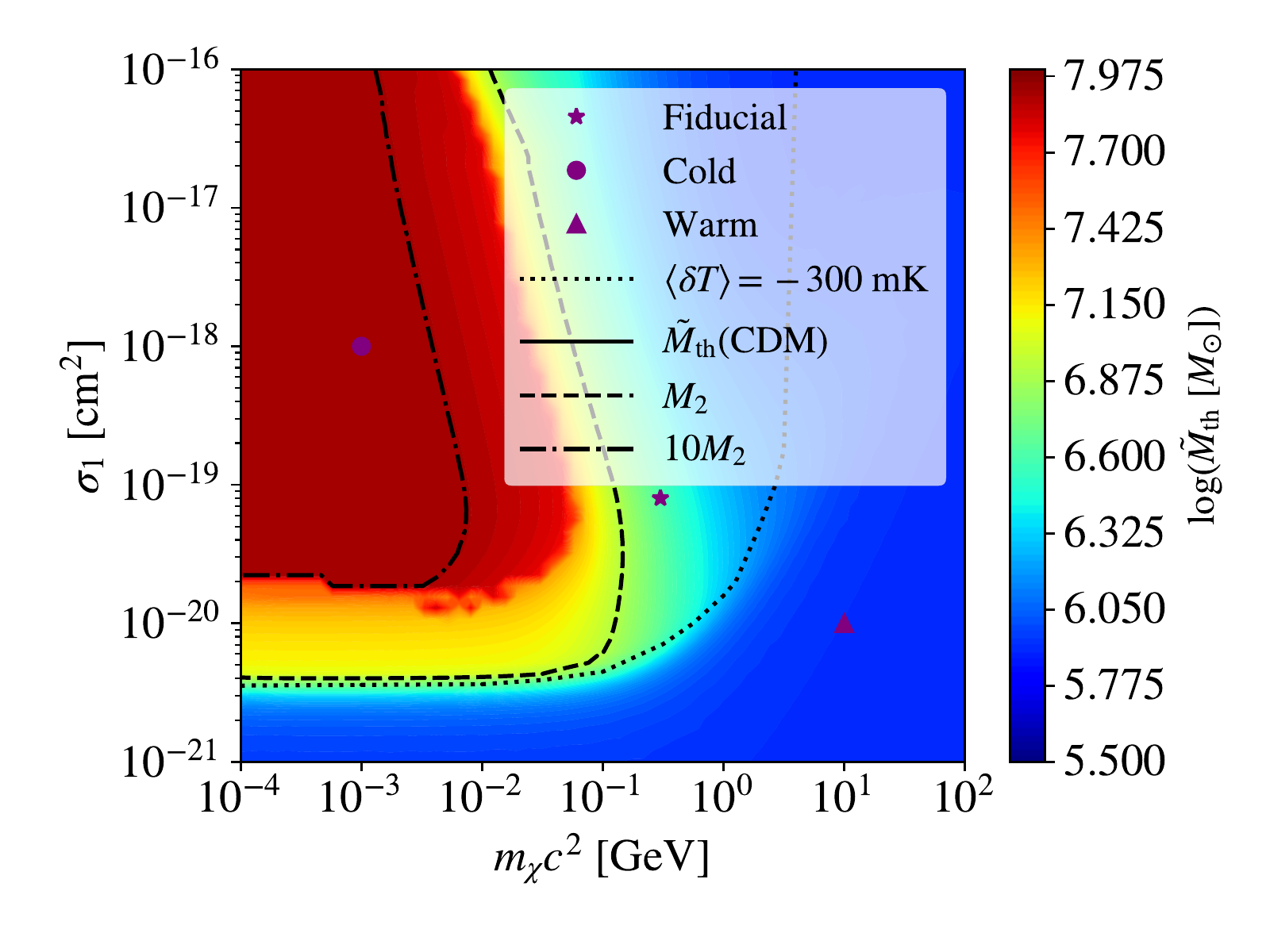}}
\subfloat[$v_{b\chi,0}=0.8\sigma_{\mathrm{rms}}$]{\includegraphics[width= 1.09\columnwidth]{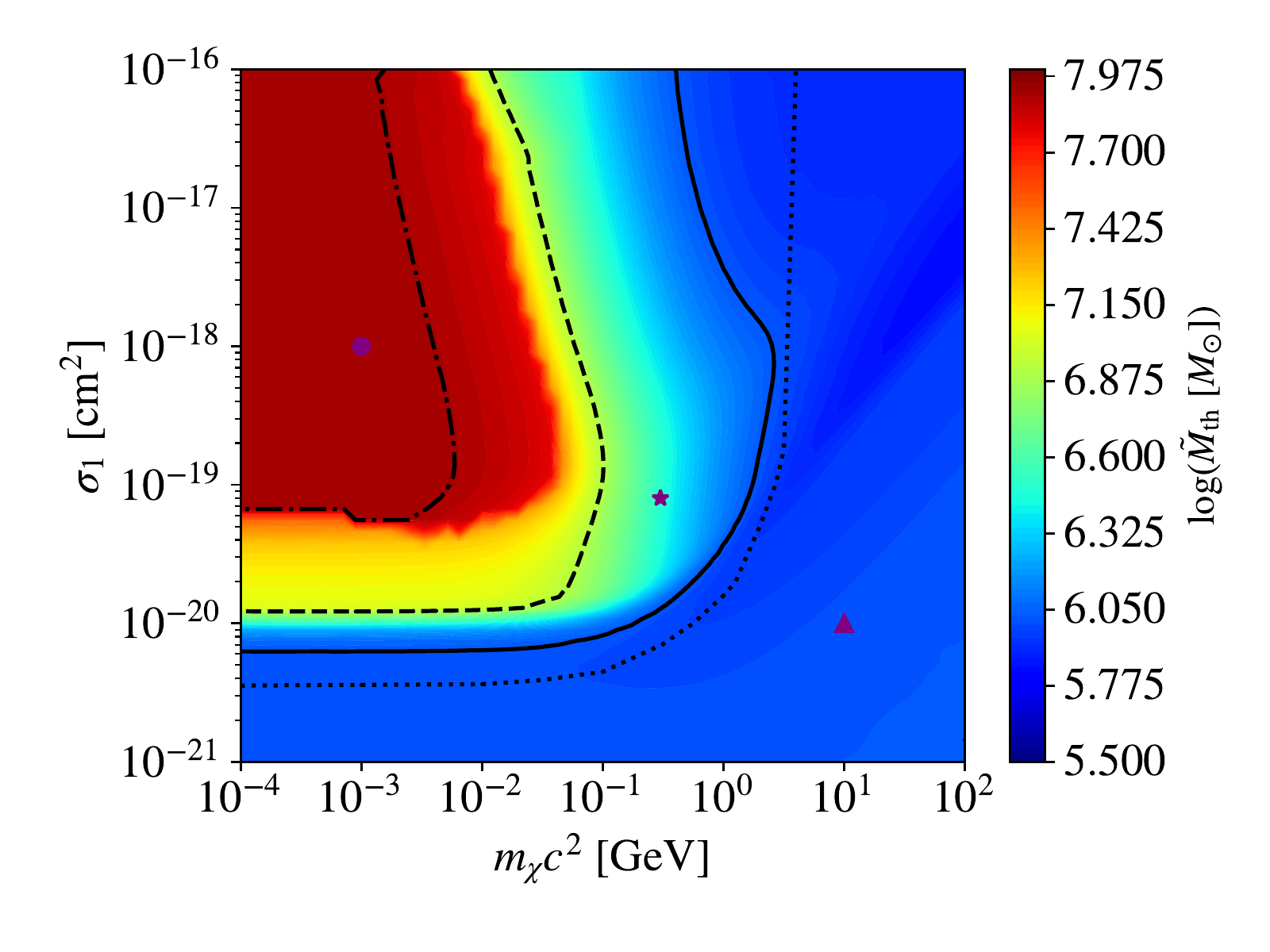}}
\vspace{-10pt}
\subfloat[$v_{b\chi,0}=1.5\sigma_{\mathrm{rms}}$]{\hspace{-11pt}\includegraphics[width= 1.09\columnwidth]{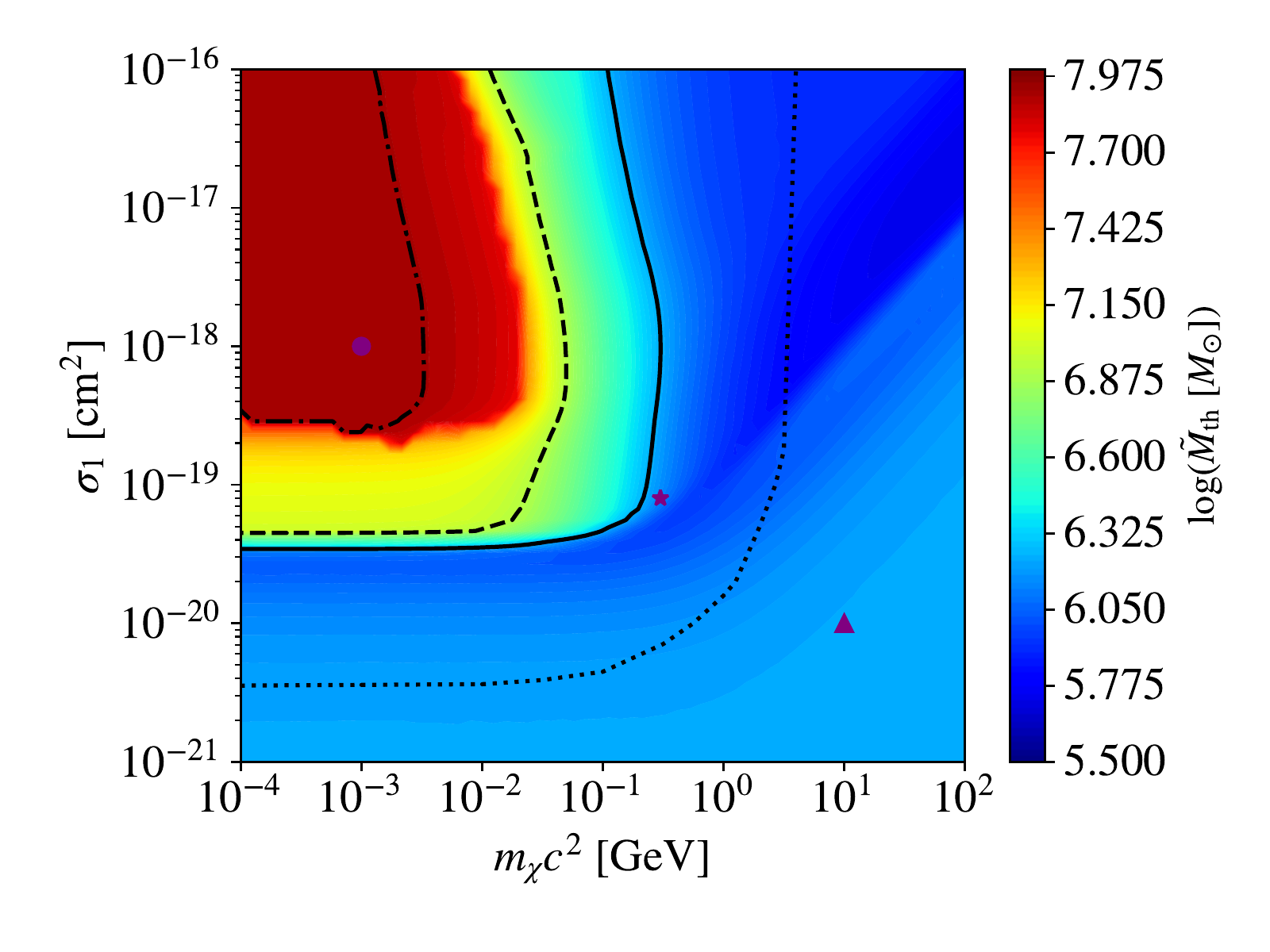}}
\subfloat[$v_{b\chi,0}=2\sigma_{\mathrm{rms}}$]{\includegraphics[width= 1.09\columnwidth]{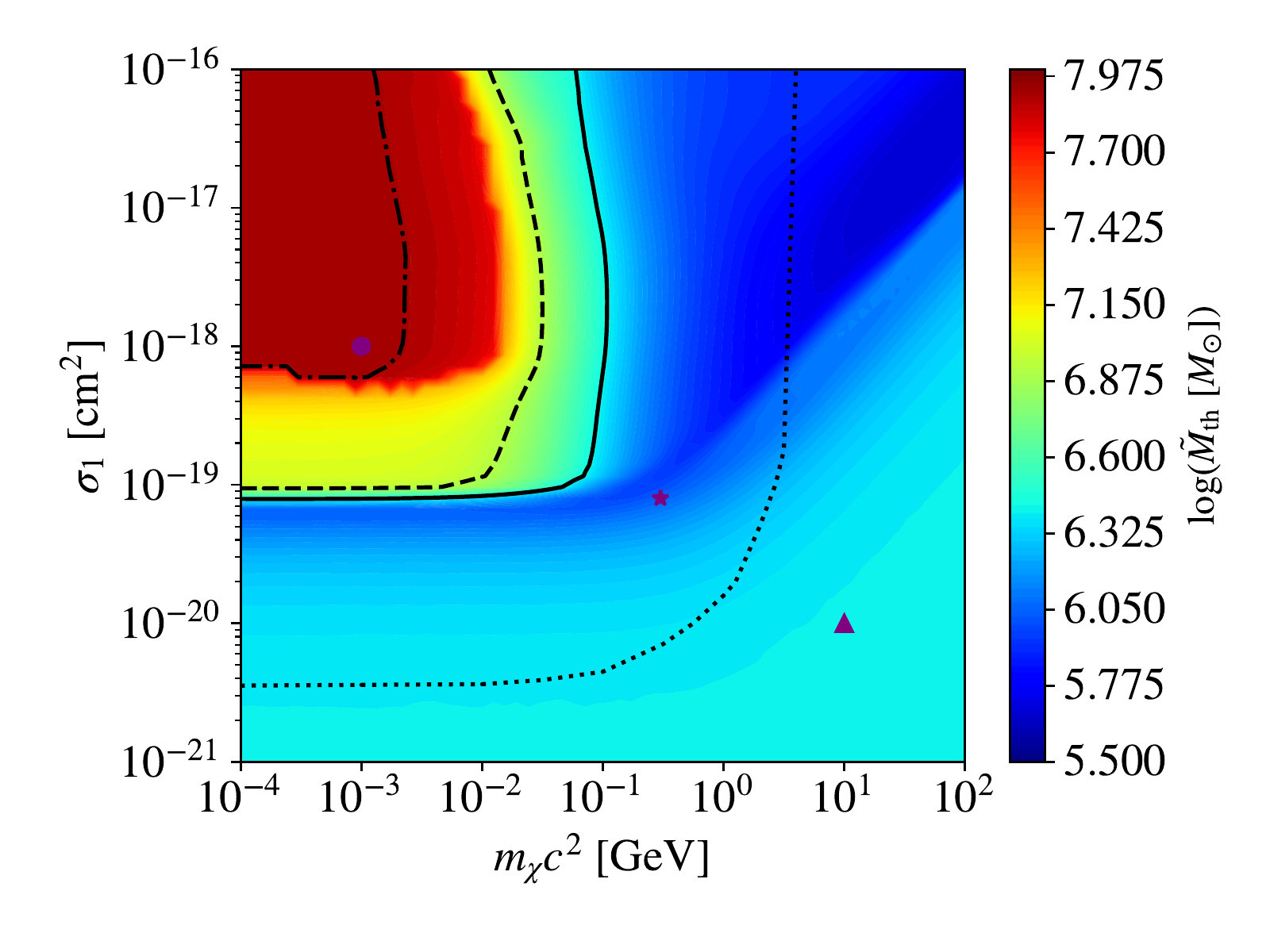}}
\vspace{-15pt}
\caption{Maps of the mass thresholds for Pop~III host haloes at $z_{\mathrm{vir}}=20$, in the $\sigma_{1}$-$m_{\chi}$ parameter space of BDMS models, for (a) $v_{b\chi,0}=0$, (b) $v_{b\chi,0}=0.8\sigma_{\mathrm{rms}}$, (c) $v_{b\chi,0}=1.5\sigma_{\mathrm{rms}}$, and (d) $v_{b\chi,0}=2\sigma_{\mathrm{rms}}$, where $\sigma_{\mathrm{rms}}=30\ \mathrm{km\ s^{-1}}$. The location of the fiducial model ($m_{\chi}c^{2}=0.3$~GeV, $\sigma_{1}=8\times 10^{-20}\ \mathrm{cm^{2}}$) is labelled with a star, the typical `cold' model ($m_{\chi}c^{2}=10^{-3}$~GeV, $\sigma_{1}=1\times 10^{-18}\ \mathrm{cm^{2}}$) with a solid circle, and the typical `warm' model ($m_{\chi}c^{2}=10$~GeV, $\sigma_{1}=1\times 10^{-20}\ \mathrm{cm^{2}}$) with a triangle. The solid contour denotes the value in the CDM model, the dashed contour the atomic cooling threshold $M_{2}$ (Equ.~\ref{e12}), and the dashed-dotted contour $10M_{2}$. The constraint based on the EDGES signal corresponding to the minimal absorption depth of $\langle \delta T\rangle=-300$~mK at 99\% confidence level \citep{nature1} is shown with the dotted contour. Here we have imposed an upper bound of $\log (M_{\max}\ [M_{\odot}]) = \log (10 M_2 \ [M_{\odot}]) \approx 7.9{}$ on the mass threshold.}
\label{f7}
\end{figure*}

\subsection{Constraining BDMS parameter space}
\label{s3.2}
We calculate the mass thresholds at $z_{\mathrm{vir}}=20$ in the BDMS parameter space with $10^{-21}\le\sigma_{1}\ [\mathrm{cm^{2}}]\le 10^{-16}$ and $10^{-4}\le m_{\chi}c^{2}\ [\mathrm{GeV}]\le 100$. 
The results are shown in Fig.~\ref{f7} for (a) $v_{b\chi,0}=0$ (b) $v_{b\chi,0}=0.8\sigma_{\mathrm{rms}}$, (c) $v_{b\chi,0}=1.5\sigma_{\mathrm{rms}}$, and (d) $v_{b\chi,0}=2\sigma_{\mathrm{rms}}$. 
A general feature is that $\tilde{M}_{\mathrm{th}}$ is significantly enhanced in the top-left corner of parameter space with low $m_{\chi}$ and high $\sigma_{1}$ (enclosed by the dashed contour), where formation of $\mathrm{H_{2}}$ is strongly suppressed in pre-virialization evolution (`cold zone'). On the other hand, if non-negligible streaming motions exist ($v_{b\chi,0}\gtrsim 0.8\sigma_{\mathrm{rms}}$), $\tilde{M}_{\mathrm{th}}$ can be somewhat reduced in the region of parameter space with high $m_{\chi}$ or low $\sigma_{1}$. Here, the effects of heating and enhanced decay of streaming motions dominate (`warm zone'). The area of the `cold zone' decreases, and that of the `warm zone' increases with $v_{b\chi,0}$, consistent with the trend found in Section~\ref{s3.1} that (frictional) heating of gas and decay of streaming motions by BDMS are more important with higher (initial) streaming velocities. 

\begin{figure}
    \vspace{-7pt}
    \hspace{-12pt}
    \includegraphics[width=1.07\columnwidth]{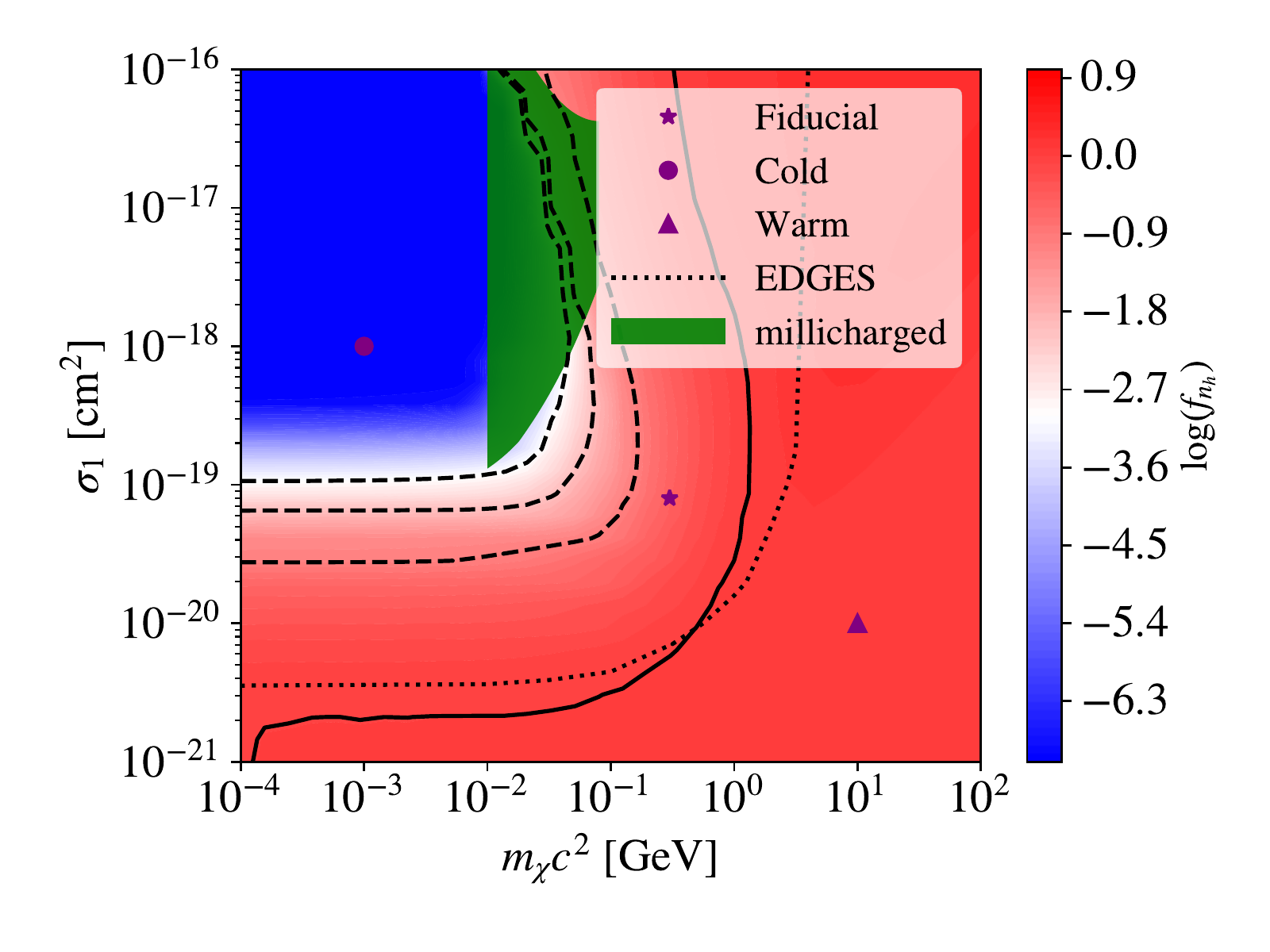}
    \vspace{-25pt}
    \caption{Map of the number density ratio, $f_{n_{h}}=\bar{n}_{h}^{\mathrm{Pop III}}(\mathrm{BDMS})/\bar{n}_{h}^{\mathrm{Pop III}}(\mathrm{CDM})$, for Pop~III hosts at $z_{\mathrm{vir}}=20$, in the $\sigma_{1}$-$m_{\chi}$ parameter space of BDMS models. The symbols follow the same convention as in Fig.~\ref{f7}. The solid contour denotes $f_{n_{h}}=1$, while the dashed ones $f_{n_{h}}=0.1,\ 0.01\ \text{and }10^{-3}$ (from bottom-right to top-left). The IGM constraint based on the EDGES signal is shown with the dotted contour \citep[][]{nature1}. For comparison, we also show the current constraints on millicharged DM from \citet{berlin2018severely}, as the green shaded region. Here, we have placed a lower bound $10^{-7}$ on $f_{n_{h}}$.}
    \label{f9}
\end{figure}

Similar to Section~\ref{s3.1}, we calculate the number density ratio $f_{n_{h}}=\bar{n}_{h}^{\mathrm{Pop III}}(\mathrm{BDMS})/\bar{n}_{h}^{\mathrm{Pop III}}(\mathrm{CDM})$ for Pop~III hosts over the same parameter space at $z_{\mathrm{vir}}=20$. Fig.~\ref{f9} shows the resulting map of $f_{n_{h}}$, where we impose a lower bound $10^{-7}$ on $f_{n_{h}}$ for clarity of presentation. 
For comparison, we also show the current constraints on millicharged DM from the CMB, light element abundances, Supernova 1987A and laboratory experiments \citep{berlin2018severely}\footnote{In \citet{berlin2018severely}, the allowed range of millicharged DM models is expressed in terms of the charge $\epsilon\sim 10^{-4}-10^{-6}$, mass fraction (within DM) $f_{\mathrm{DM}}\sim 0.003-0.02$ and mass $m_{\chi}c^{2}\sim 10-80$~MeV of millicharged DM particles. We map this range to our phenomenological parameter space based on Formula (1) from \citet{munoz2018insights}, and the fact that the heat-exchange rate $\dot{Q}_{b}\propto f_{\mathrm{DM}}\sigma_{1}u_{\mathrm{th}}^{-3}\propto f_{\mathrm{DM}}\sigma_{1} T_{\mathrm{fin}}^{-3/2} $ at thermal equilibrium, where $T_{\mathrm{fin}}\simeq T_{b}(\mathrm{CDM})/[1+f_{\mathrm{DM}}(6\ \mathrm{GeV}/m_{\chi}c^{2})]$.}, as the green shaded region with $10^{-2}\lesssim m_{\chi}c^{2}\ [\mathrm{GeV}]\lesssim 8\times 10^{-2}$ and $\sigma_{1}\sim 10^{-19}-10^{-16}\ \mathrm{cm^{2}}$. 
It turns out that the region with $\sigma_{1}\gtrsim 10^{-19}\ \mathrm{cm^{2}}$ and $m_{\chi}c^{2}\lesssim 3\times 10^{-2}$~GeV is effectively ruled out, since there $f_{n_{h}}\lesssim 10^{-3}$, implying that Pop~III star formation is significantly suppressed. These constraints from the perspective of the non-linear regime of structure formation nicely complement those based on the 21-cm absorption signal from the IGM background (see Figure~3 in \citealt{nature1}). 
In the specific case of millicharged DM, our results further rule out about half of the allowed region in parameter space. Besides, in most parts of the remaining allowed region, $f_{n_{h}}\lesssim 0.1$, implying that the star formation efficiency of Pop~III stars required for millicharged DM to explain the observed 21-cm signal should be higher than that in the CDM case by a factor of 10.

In this study, we focus on BDMS models with a $v^{-4}$ velocity dependence. For models with other velocity dependencies, the trends will be similar in the relevant parameter spaces (see Fig.~\ref{f7} and \ref{f9}). That is to say, at a fixed (virialization) redshift, smaller DM particle masses and larger cross-sections lead to stronger cooling, lower $\mathrm{H}_{2}$ abundances and, thus, higher mass thresholds for star formation. But the detailed thermal histories will be different. In general, if the velocity dependence follows a power law $\sigma\propto v^{n}$, the larger the power-law index $n$, the earlier the effect of BDMS becomes significant, because baryons and DM are hotter for larger encounter velocities at higher redshifts (see FIG.~2 of \citealt{dvorkin2014constraining}). In the context of the EDGES signal, to achieve the same (sufficiently low) IGM temperature at $z\sim 20$, efficient cooling (heating) of baryons (DM) tends to occur earlier for larger $n$, which is more likely to violate CMB observations. 

\section{Discussions and conclusions}
\label{s4}
We derive the mass thresholds $\tilde{M}_{\mathrm{th}}$ of dark matter (DM) haloes in which primordial gas can cool efficiently to form Population III (Pop~III) stars, under baryon-dark-matter scattering (BDMS), by calculating the relevant thermal and chemical histories with a one-zone model. We focus on the BDMS model with interaction cross-section $\sigma=\sigma_{1}[v/(1\ \mathrm{km\ s^{-1}})]^{-4}$, where $v$ is the relative velocity of the encounter. 
We assume that the velocity distribution of DM particles can be approximated with the Maxwell-Boltzmann distribution, which is only valid with sufficient DM self-interactions. For weakly self-interacting DM, we expect the ideal gas approximation to result in uncertainties in the energy-/momentum-transfer rate of a factor of a few, based on the analysis in \citet{ali2019boltzmann}.

In our calculation, we have taken into account the effect of streaming motion between DM and gas, with a parameterized model calibrated to simulation results (see Sec.~\ref{s2.3} for details). However, we do not consider the effect of Lyman-Werner (LW) photons, which can enhance the mass threshold. We expect that the constraints on BDMS summarized below will be tightened when the LW field is included. Besides, we only model the thermal legacies of virialization without a full treatment of the relevant dynamical effects (e.g., ionization associated with shocks), which means that our model is only suitable to Pop~III star formation. 

Despite these caveats, our results provide new constraints on BDMS models from the non-linear regime of early structure formation:
\begin{itemize}
\item In the fiducial model of BDMS with $m_{\chi}c^{2}=0.3$~GeV and $\sigma_{1}=8\times 10^{-20}\ \mathrm{cm^{2}}$ that can accommodate the 21-cm absorption depth measured by EDGES \citep{nature}, the mass threshold is enhance by a factor of a few, and the cosmic average number density of Pop~III hosts is reduced by a factor of $3-10$, in the EDGES epoch $z\sim 15- 20$, compared with the case of the standard $\Lambda$CDM model. Therefore, if this model is to self-consistently explain both the strength and timing of the observed 21-cm signal, differences from the CDM model in astrophysical parameters (e.g., star formation efficiency of Pop~III stars) must exist.
\item The region with $\sigma_{1}\gtrsim 10^{-19}\ \mathrm{cm^{2}}$ and $m_{\chi}c^{2}\lesssim 3\times 10^{-2}$~GeV in the BDMS parameter space is ruled out. In this region, the number density of Pop~III hosts is reduced by at least three orders of magnitude at $z\sim 20$ compared with the case of CDM, indicating that formation of Pop~III stars is significantly suppressed for $z>20$, inconsistent with the timing of the observed 21-cm absorption signal. These constraints complement those based on the 21-cm absorption signal from the IGM background \citep{nature1}. The remaining allowed region in the BDMS parameter space is a `belt' with $3\times 10^{-2}\lesssim m_{\chi}c^{2}\ \mathrm{[GeV]}\lesssim 0.6$ and $4\times 10^{-21}\lesssim \sigma_{1}\ [\mathrm{cm^{2}}]\lesssim 10^{-19}$. For the specific case of millicharged DM, our results further tighten the existing constraints from \citet{berlin2018severely} by ruling out the models with $m_{\chi}c^{2}\sim 10-40$~MeV.
\end{itemize}

Considering the exploratory nature of this semi-analytical work, it is important to follow up with cosmological hydrodynamic simulations for BDMS models in the future. With simulations, one can include more physics (e.g., stellar feedback and radiation fields), and study early structure formation in greater detail (e.g., internal structures of DM haloes and star-forming clouds, star formation and chemical enrichment histories, statistical and global radiation signature such as UV luminosity function and cosmic radio background), as has already been done for CDM, WDM, and fuzzy DM models (e.g., \citealt{yoshida2003early,oshea2006,gao2007lighting,dayal2017reionization,hirano2017first,jaacks2018legacy,jaacks2018baseline,boyuan2019}). However, this is not trivial, requiring new numerical techniques of sampling the phase space of DM with simulation particles and implementation of the scattering processes. BDMS models would deserve such efforts even more if the EDGES signal were confirmed with follow-up measurements, accentuating the possible tension between the standard $\Lambda$CDM model and 21-cm observations. At the current stage, semi-analytical models (e.g., \citealt{madau18,Mirocha2018,Anna2019}) are important to explore all possible implications from the 21-cm absorption signal on early structure formation. Along this direction, it is also interesting to extend this work to other interpretations of the EDGES signal beyond BDMS, such as the early excess radio background \citep{feng2018enhanced,fraser2018edges,Mirocha2018,boyuan2019}. The role of early cosmological structure formation as a precision laboratory for DM particle physics is likely to further grow in the next decade.

\section*{Acknowledgements}
This work was supported by National Science Foundation (NSF) grant AST-1413501. 
Support for this work was provided by NASA through the NASA Hubble Fellowship grant HST-HF2-51418.001-A awarded by the Space Telescope Science Institute, which is operated by the Association of Universities for Research in  Astronomy, Inc., for NASA, under contract NAS5-26555. 

\bibliographystyle{mnras}
\bibliography{ref} 

\label{lastpage}
\end{document}